\documentclass[12pt]{article}
\usepackage[margin=1.in]{geometry}

\usepackage{authblk, amsthm, amsmath, amssymb, bm,enumerate, mathrsfs,mathtools, enumitem}
\usepackage{latexsym,color,verbatim, array, multirow, bigints}
\usepackage{multirow}
\usepackage{array, tabularx}
\usepackage{subfigure}

\usepackage{etoolbox}
\makeatletter
\patchcmd{\@makecaption}
  {\parbox}
  {\advance\@tempdima-\fontdimen2} 
  {}{}
\makeatother  
\usepackage[caption = false]{subfig} 
\usepackage{enumerate}
\usepackage{comment}
\usepackage{natbib}

\usepackage{float}
\usepackage{graphics,amssymb,color}
\usepackage{graphicx,booktabs}
\usepackage{grffile}
\usepackage{mathtools}
\usepackage{algorithm}
\usepackage[noend]{algpseudocode}
\usepackage{xfrac}

\usepackage{natbib}

\newcommand{\real}{\mathbb{R}}

\newtheorem{theorem}{Theorem}

\newtheorem{remark}{Remark}

\newcommand{\grad}{\nabla}

\newcommand{\Beta}{\boldsymbol\beta}
\newcommand{\Alpha}{\boldsymbol\alpha}
\newcommand{\Zeta}{\boldsymbol\zeta}
\newcommand{\Mu}{\boldsymbol\mu}

\makeatletter
\renewcommand*\env@matrix[1][\arraystretch]{%
  \edef\arraystretch{#1}%
  \hskip -\arraycolsep
  \let\@ifnextchar\new@ifnextchar
  \array{*\c@MaxMatrixCols c}}
\makeatother

\newcommand\MyBox[2]{
  \fbox{\lower0.75cm
    \vbox to .6cm{\vfil
      \hbox to 1cm{\hfil\parbox{0.8cm}{#1\\#2}\hfil}
      \vfil}%
  }%
}
\usepackage{lipsum}

\usepackage{xr-hyper}

\newcolumntype{M}[1]{>{\centering\arraybackslash}m{#1}}
\newcolumntype{N}{@{}m{0pt}@{}}

\newtheorem{Proposition}{Proposition}

\usepackage[normalem]{ulem}

\begin{document}
\title{Integrative Bayesian models using Post-selective Inference: a case study in Radiogenomics }
\date{}

\author[1]{Snigdha Panigrahi\thanks{psnigdha@umich.edu}}
\author[2,3]{Shariq Mohammed}
\author[2,3,4,5]{Arvind Rao}
\author[2,3]{Veerabhadran Baladandayuthapani\thanks{veerab@umich.edu}}

\affil[1]{Department of Statistics, University of Michigan.}
\affil[2]{Department of Biostatistics, University of Michigan.}
\affil[3]{Department of Computational Medicine and Bioinformatics,\\ University of Michigan.}
\affil[4]{Department of Biomedical Engineering, University of Michigan.}
\affil[5]{Department of Radiation Oncology, University of Michigan.}

\maketitle

\begin{abstract}
Integrative analyses based on statistically relevant associations between genomics and a wealth of intermediary phenotypes (such as imaging) provide vital insights into their clinical relevance in terms of the disease mechanisms.
Estimates for uncertainty in the resulting integrative models are however unreliable unless inference
 accounts for the selection of these associations with accuracy.
In this article, we develop selection-aware Bayesian methods which:
 (i)  counteract the impact of model selection bias through a ``selection-aware posterior" in a flexible class of integrative Bayesian models post a selection of promising variables via $\ell_1$-regularized algorithms;
 (ii) strike an inevitable tradeoff between the quality of model selection and inferential power when the same dataset is used for both selection and uncertainty estimation. 
Central to our methodological development, a carefully constructed conditional likelihood function deployed with a reparameterization mapping provides notably tractable updates when gradient-based MCMC sampling is used for estimating uncertainties from the selection-aware posterior.
Applying our methods to a radiogenomic analysis, we successfully recover several important gene pathways and estimate uncertainties for their associations with patient survival times.
\end{abstract}

\section{Introduction}
Our methodology in the present paper is motivated by a radiogenomic analysis in low-grade gliomas (LGG), a type of brain cancers.
Briefly, a radiogenomic analysis ascertains associations between imaging outcomes obtained from radiological imaging modalities, e.g. magnetic resonance imaging (MRI) with molecular and genomic markers. 
We pursue post-selective inference, alternately called ``selection-aware" inference, in a two-stage integrative modeling framework built on a sequential flow of information: genomics to imaging to clinical outcomes. In the first stage, we glean important genomic variables which help us identify the variables associated with (multiple) imaging outcomes, called radiogenomic variables \citep{zhang2019radio}. The second stage then assesses the clinical relevance of these radiogenomic variables on clinical outcomes. In some sense, this follows the natural progression of cancer, where genomic changes initiate tumor formation and development, that are subsequently assessed using imaging, and finally manifest clinical outcomes (e.g. survival) are assessed as the eventual clinical phenotypes. 

There exist fundamental gaps between the use of integrative models based on the selected associations between different modalities of information and reliable estimation of uncertainties for the matched parameters.
To bridge the void, we recognize the call for ``selection-aware inference" in order to systematically counteract the bias incurred in the selection of integrative models \citep{benjamini2005false, berk2013valid, exact_Lasso}.
Intertwined with this goal, we notice an inevitable tradeoff of information between the quality of model selection and inferential power for uncertainty estimates from these models. 
In all realistic scenarios, the extent to which this balance is accomplished has severe implications
on the number of discoveries and the power of making these discoveries.
Impaired by the high dimensional nature of genomic variables, known to share complex correlation structures and sparse in terms of their associations with the outcomes, and by the availability of (relatively) small sample sizes,
the repercussions of unreliable inference and low power of discoveries can be quite profound in a radiogenomic case study. 
Our  Bayesian methods in the paper demonstrate the potential of reusing samples towards two indispensable goals of inference: (i) 
an effective integrative modeling of clinical outcomes with interpretable parameters in terms of their mechanisms, (ii) a significant reduction in the variance of the matched estimates
for the parameters within the integrative models while overcoming the hazardous effects of selection bias at the same time.
Before proceeding further, we provide an overview of our methodological development through a schematic snapshot of the integrative pipeline for inference and draw connections with related literature.

\section{Schematic overview and related literature}
\label{schematic:related}

\textbf{Overview.} \ \ Introducing some basic notations, we denote the outcome variable, the matrix of $p$ explanatory variables and the matrix of $L$ intermediary outcomes, all measured across the same set of $n$ samples, by $\boldsymbol{\mathrm{y}}\in \real^n$, $\mathbf{G}\in \real^{n\times p}$ and $\boldsymbol{\mathcal{I}}\in \real^{n\times L}$ respectively. These measurements represent in the radiogenomic case study the clinical outcome, the genomic variables and the imaging outcomes respectively.
We let $\boldsymbol{\mathcal{I}}_l$ stand for the imaging outcome $l$, which is the $l$-th 
 column of $\boldsymbol{\mathcal{I}}$, and let $\mathbf{G}_F$ represent the submatrix of $\mathbf{G}$ containing the subset of columns indexed by $F\subset \{1,2, \cdots, p\}$.  
Through the paper, we use the notation $\rho(\mathbf{d}; \mathbf{\Beta}_E, \mathbf{\Sigma}_E)$ for a normal density function with mean and covariance $\mathbf{\Beta}_E$ and $\mathbf{\Sigma}_E$ respectively, evaluated at $\mathbf{d}$, $\text{diag}(\mathbf{V})$ for a diagonal matrix with the vector $\mathbf{V}$ along the diagonal, $\text{vec}(V_1, \dots, V_k)$ to denote a vector with the entries $V_j$, $j\in \{1,\dots, k\}$ and use $[\mathbf{A}]_j$ to denote the $j$-th column vector of the matrix $\mathbf{A}$ wherever needed.

\begin{figure}
  \centering
  \begin{tabular}{@{}c@{}}
    \includegraphics[height=180pt,width=0.90\linewidth]{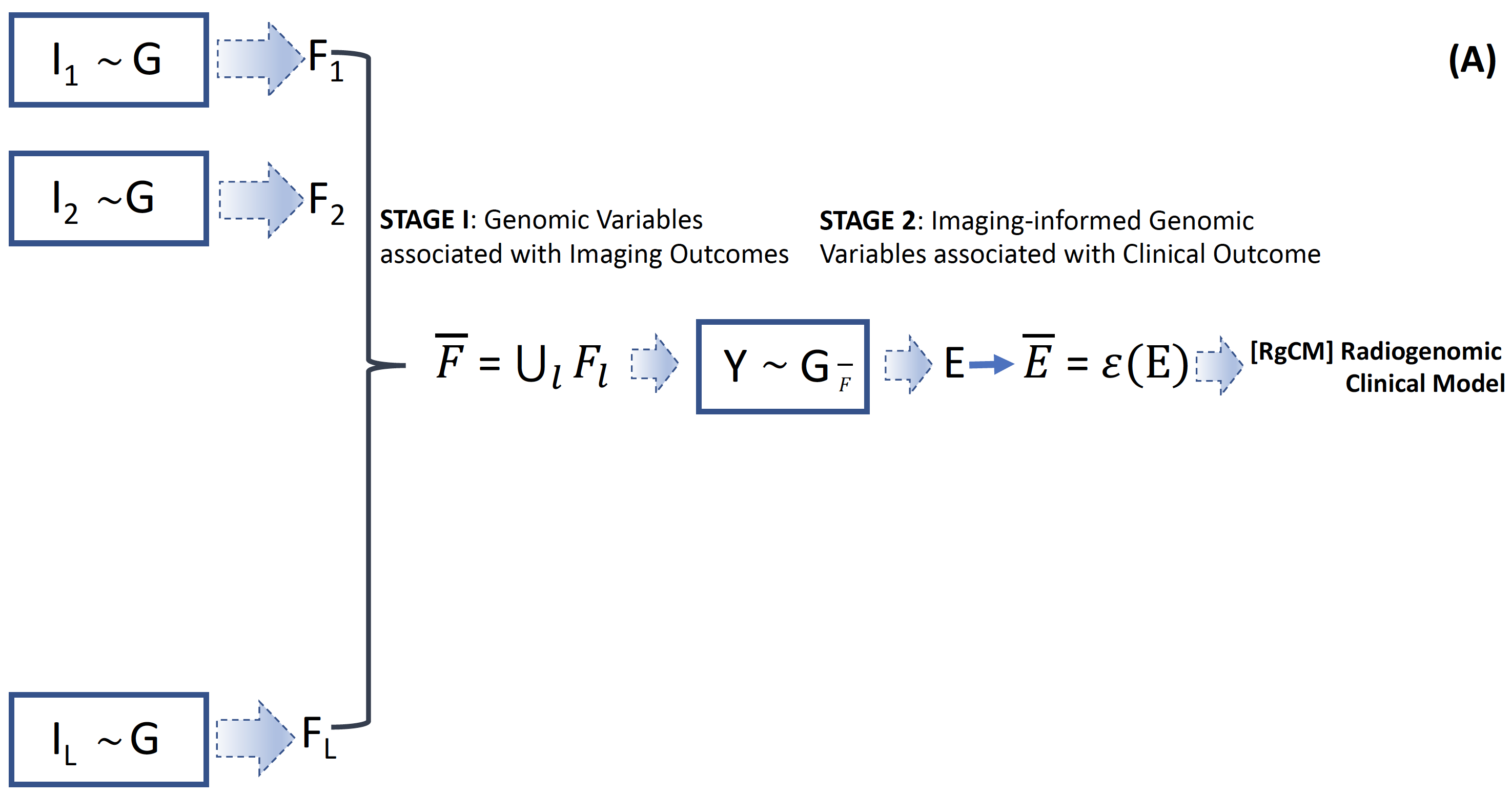} \\ [\abovecaptionskip]
      \end{tabular}
  
  \vspace{\floatsep}
  
  \begin{tabular}{@{}c@{}}
    \includegraphics[height=130pt,width=0.91\linewidth]{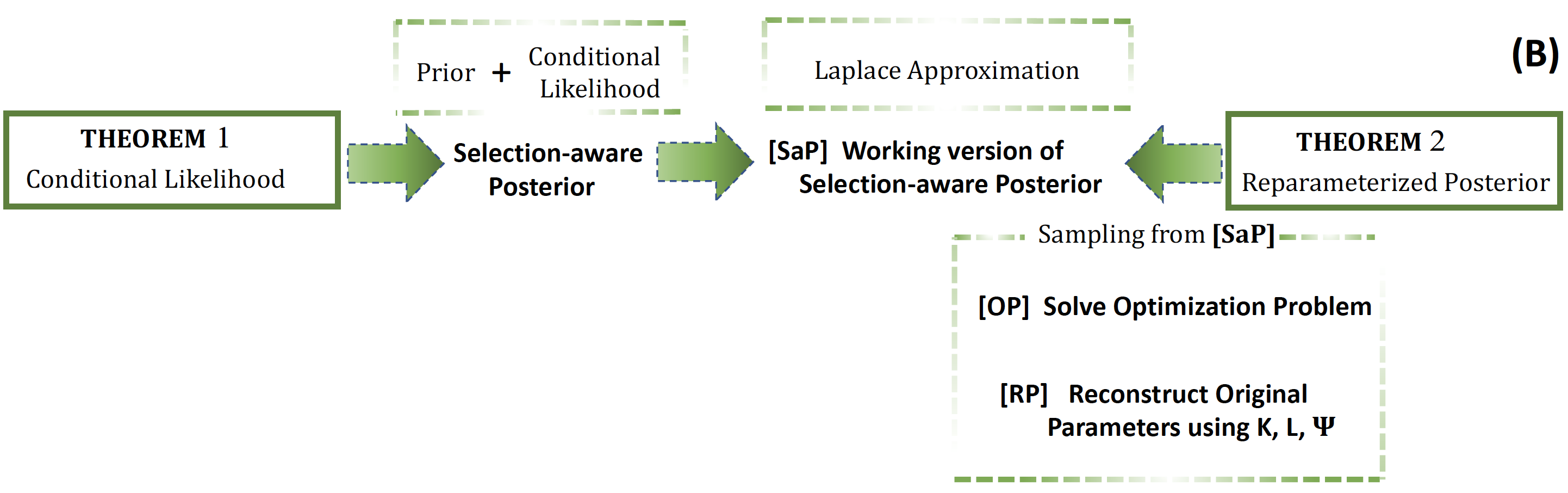} \\ [\abovecaptionskip]
  \end{tabular}
  \caption{Panel \textbf{(A)}. \ \ \underline{Schematic representation of the inputs and outputs of the} \underline{ selection pipeline divided into two stages}: The first stage gives a candidate set of genomic variables associated with the intermediary imaging phenotypes; the second stage selects promising variables from an imaging-informed set of genomic variables (from the first stage) that are statistically associated with the clinical outcome. 
   \hspace{\textwidth}
  Panel \textbf{(B)}. \ \ \underline{Schematic representation of the methodological development}: Theorem 1 gives a simplified expression for the conditional likelihood function which in conjunction with the prior post selection leads us to a selection-aware posterior.
Theorem 2 provides us tractable optimization-based updates [\textbf{OU}]  to sample from a working version of the selection-aware posterior [\textbf{SaP}] through a reparameterization mapping [\textbf{RP}].
Incurring no additional cost, we construct from the reparameterization mapping samples for our original target, the parameters within the integrative radiogenomic clinical model [\textbf{RgCM}].}
 \label{schematic:pipeline}
\end{figure}

Figure \ref{schematic:pipeline}(A) depicts how the selection steps inform our integrative models, detailed out in Section \ref{motivation:model}.
Divided into two stages, we deploy a multiple regression framework in the first step of this pipeline in order to select genomic variables associated with at least one of the intermediary imaging outcomes.
We index the selected set of genomic variables by $\bar{F} \subset\{1,2,\cdots, p\}$. 
In the second step, we select from $\bar{F}$ the variables that are further associated with the clinical outcome. 
Calling this set $E$ with cardinality $|E|$, the output at this stage results in an integrative radiogenomic clinical model [\textbf{RgCM}], using $\mathcal{E}(E)= \bar{E}\subset\{1,2,\cdots, p\}$, where $\mathcal{E}$ is a deterministic mapping that takes $E$ as the input and returns $\bar{E}$.

Figure \ref{schematic:pipeline}(B) outlines the two core inferential results we establish in the paper to validate the use of integrative models from the pipeline in Figure \ref{schematic:pipeline}(A) through a ``selection-aware posterior". 
Deferring the technical details to Section \ref{methods}, in a nutshell, this selection-aware posterior uses a conditional likelihood, obtained by conditioning out the observed event wherein the pipeline selects the set of genomic variables indexed by $E$, jointly with a prior post selection. 
To this end, Theorem \ref{correction:factor:cond} identifies a simplified expression for the conditional likelihood function, in the sense that the truncation region associated with the conditioning event can be very simply expressed in terms of sign constraints on the data variables.
Theorem \ref{reparam:map} then enables us tractable optimization-based updates [\textbf{OP}]  to sample from a working version of the selection-aware posterior [\textbf{SaP}] through a reparameterization mapping [\textbf{RP}].
Incurring no additional cost, we construct from the reparameterization mapping samples for our original target, the parameters within the radiogenomic clinical model [\textbf{RgCM}].

\noindent \textbf{Related work.} \ \ Drawing connections with existing work, our approach of using a ``selection-aware posterior" is anchored within a conditional proposal for Bayesian models post selection \citep{yekutieli2012adjusted, selective_bayesian}.
 Such an approach deploys conditioning through the likelihood to discard the information from data consumed in model selection. 
Adopting the Bayesian perspective as opposed to a frequentist solution to the post-selective problem  \citep[][among others]{exact_Lasso, randomized_response, lee_screening, panigrahi2018selection}
admits several flexibilities for subsequent estimation of uncertainties. 
The latter line of work outlines inference for real-valued parameters after selection, in models where the outcome mean is simply modeled as $\mu\in \mathbb{R}^n$ without specifying a relation with the explanatory variables. 
Our Bayesian prescription on the other hand permits a joint estimation of vector-valued parameters and functions thereof in a  flexible class of models, including for example models based on an interplay between the output of the selection pipeline and prior knowledge for the explanatory variables (see Remark \ref{rem:1}).
Especially noteworthy, the present methods allow us to take full advantages of a Bayesian implementation without losing the computing efficiency of a selection-aware frequentist solution.

As a prelude to the technical development, let $\Beta_{\bar{E}}$ be the parameters in the radiogenomic clinical [\textbf{RgCM}].
Our starting point for inference is the selection-aware posterior for $\Beta_{\bar{E}}$, proportional to:
$$ \Big\{\mathbb{P}[\; \mathcal{E}_0  \;\lvert \;\Beta_{\bar{E}}]\Big\}^{-1} \cdot \pi(\mathbf{D};\Beta_{\bar{E}}),$$ 
where $\mathbb{P}[\; \mathcal{E}_0  \;\lvert \;\Beta_{\bar{E}}]$ is the probability for the event of selection $\mathcal{E}_0$, the set of all realizations of data leading to the selected set $E$, $\pi(\mathbf{D};\Beta_{\bar{E}})$ is the usual (ignoring the effects for selection) posterior based on observed data $\mathbf{D}$.
Previous proposals \citep{selective_bayesian, panigrahi2018scalable} establish a statistically consistent approximation for the probability of selection, enabling a working version of the otherwise intractable selection-aware posterior. 
Sampling from the working posterior however remains largely an arduous task in the high dimensional regime, due to impediments from both geometric and analytic angles.
Geometrically, the conditional likelihood is truncated to an event of selection characterized by a union of polyhedral regions that is not easily amenable to sampling for inference. 
From the analytic perspective, several variables must be integrated out to calculate the probability of selection that sets apart our posterior from the analog ignoring bias from selection.

Simplifying the complex geometry of the truncation region and bypassing intensive integrations, our new methods in the paper quite remarkably facilitate very tractable updates for gradient-based sampling from the selection-aware posterior.
At the core of every update, we solve only an $|E|$ dimensional convex optimization.
A substantial leap forward from existing proposals, a reparameterization mapping applied to a carefully constructed conditional likelihood function reduces the effective dimension of inferential updates by orders of magnitude smaller than the size of the high dimensional, initial set of $p$ explanatory variables. 
In doing so, the tradeoff our selection-aware Bayesian methods strike  between selection and inference stands in stark contrast with 
sample splitting \citep[][for example]{hurvich1990impact}, an appealing tool of choice for practitioners to negate selection bias.
With sample sizes as small as $60$, ignoring a fraction of the samples for either of the two goals is highly suboptimal for integrative inference.
Our numerical experiments illustrate this tradeoff of information and highlight the advantages our methods enjoy over splitting at different resolutions in terms of the support recovery of models and the power of their uncertainty estimates.
We note that a more common Bayesian prescription approaches variable selection and inference in a single shot through shrinkage priors \citep[][among others]{george1997approaches, mitchell1988bayesian, park2008bayesian}, modeling the data before selection.
Our methods on the other hand enable inference for a selection-aware model, and are based on the view that selection, in the presence of sparsity, can be harnessed to decide models with fewer and more interpretable parameters, in terms of their mechanisms.

We structure the remaining paper as follows. Section \ref{motivation:model} outlines our modeling framework after we
state the selection algorithms resulting in our integrative models. 
Section \ref{methods} discusses a selection-aware posterior and develops a tractable Bayesian framework amenable for gradient-based sampling from this posterior. 
Section \ref{simulation} explores the potential of our proposal by simulating integrative models using the actual genomic and imaging measurements from a radiogenomic case study. 
Section \ref{real:data} furnishes uncertainty estimates for the effects of biologically relevant gene pathways which we recover after utilizing the associations of the pathways with the imaging and clinical outcomes for LGG.
\vspace{-5mm}

\section{Modeling framework}
\label{motivation:model}

Adopting a two-stage selection for integrative modeling, we solve $L$ LASSO regression problems \citep{tibs_lasso} to select the promising explanatory variables associated with $\boldsymbol{\mathcal{I}}_l$, an intermediary (imaging) outcome:
\begin{equation}
\label{first:stage:Lasso}
(\widehat{\Alpha}_{F_l}, \mathbf{0}) = \arg\min_{\Alpha} \frac{1}{2}\|\boldsymbol{\mathcal{I}}_l - \mathbf{G}\Alpha\|^2_2 + \lambda_l \|\Alpha\|_1 \text{ for } l\in \{1,2,\cdots, L\};
\end{equation}
$\lambda_l$ is the tuning parameter for the $\ell_1$ penalty.  
Fixing $\bar{F}= \cup_{l=1}^L F_l$ with cardinality $|\bar{F}|$, $\bar{F}$ gives us a candidate set of explanatory genomic variables such that each variable is associated with at least one of the intermediary phenotypic outcomes.

We follow \eqref{first:stage:Lasso} with a randomized version of the LASSO regression \citep{randomized_response, panigrahi2018selection, panigrahi2018scalable, harris2016selective} to introduce a tradeoff between the quality of model selection and inferential power.
This strategy perturbs the canonical algorithm with a randomization variable $\boldsymbol{\mathrm{r}}\sim N(0, \eta^2 \cdot \mathbf{I})$ 
independent of $\boldsymbol{\mathrm{y}}$ and $\boldsymbol{\mathcal{I}}_l, \ l\in \{1,2\cdots, L\}$ to solve:
\begin{equation}
\label{second:stage:Lasso}
(\widehat{\Beta}_{E}^{\;\text{LASSO}}, \mathbf{0}) = \arg\min_{\Beta} \frac{1}{2}\|\boldsymbol{\mathrm{y}} - \mathbf{G}_{\bar{F}}\Beta\|^2_2 +  \|\mathbf{\Lambda} \Beta\|_1 +\frac{ \epsilon}{2} \cdot \|\Beta\|^2 -\boldsymbol{\mathrm{r}}^T \Beta.
\end{equation} 
In \eqref{second:stage:Lasso}, $\mathbf{\Lambda}= \text{diag}(\lambda_1, \cdots, \lambda_{\bar{F}})$, the diagonal entries are $\ell_1$-penalty weights for the $|\bar{F}|$ variables we select as significant associations with the intermediary outcomes.
Fixing these weights to be inversely proportional to the number of times an explanatory variable is 
selected across the $L$ regularized queries in the previous step is one such concrete way to incorporate the relative importance of each variable in \eqref{second:stage:Lasso}.
We identify $E$ as the set of nonzero LASSO estimates, giving us the set the explanatory variables statistically associated with both the intermediary and clinical outcomes. 
 Noticeably, the optimization objective \eqref{second:stage:Lasso} differs from a canonical version of LASSO through an additional term, 
which is linear in the randomization instance $\boldsymbol{\mathrm{r}}$ and an $\ell_2$ penalty with a small positive coefficient $\epsilon>0$. 
We use a small value for the $\ell_2$ penalty in the objective to ensure the existence of a solution for the randomized problem. 
Consistent with an elastic net type-penalty, this choice can be readily generalized to accommodate a non-trivial version of the ridge penalty.

Turning our attention to the integrative model post selection, we define:
\begin{equation}
\label{mapping}
\mathcal{E}: E \to \bar{E},
\end{equation}
a mapping applied to $E$ that returns $\bar{E}$, a subset of the $p$ explanatory variables, with cardinality $|\bar{E}|=q$. 
Specifically, \eqref{mapping} allows us the flexibility to incorporate interactions between preexisting knowledge about the explanatory variables with the output 
of the two-stage pipeline in \eqref{first:stage:Lasso} and \eqref{second:stage:Lasso}. 
For instance, adding variables to the selected set $E$ that might have been missed in the $\ell_1$ regularized selection steps, based on pathway annotations or previously validated clinical analyses are examples of some practical choices for this mapping.
Then, we assume a linear dependence between $\boldsymbol{\mathrm{y}}$ and $\mathbf{G}_{\bar{E}}$ under a fixed predictor matrix framework. Letting $\mathbf{G}^T_{i,\bar{E}}$ denote the $i$-th row of the matrix $\mathbf{G}_{\bar{E}}$, each sample $i$ is identically and independently distributed as follows:
\begin{equation}
\label{model:blackbox}
\begin{aligned}
& [\text{\textbf{RgCM}}]: \ \ \ \mathrm{y}_{i} = \mathbf{G}^T_{i,\bar{E}} \Beta_{\bar{E}} + \epsilon_i, \text{ where } \epsilon_i \sim N(0, \sigma^2).
\end{aligned}
\end{equation}
This is our primary outcome model of interest. 

For modeling the associations between the intermediary outcomes and explanatory variables, we assume for now
\begin{equation}
\label{model:intermediary}
\mathcal{I}_{i, l} = \mathbf{G}^T_{i; F_l} \Alpha_{F_l} + \Psi_{i,l}, \; l\in \{1,2,\cdots, L\}, \  \mathbf{\Psi}_{i}= (\Psi_{i,1}, \cdots, \Psi_{i,L}) \stackrel{i.i.d.}{\sim} N (0, \mathbf\Sigma_I);
\end{equation}
$\mathcal{I}_{i, l}$ is a sample for the intermediary outcome $\boldsymbol{\mathcal{I}}_{l}$ indexed by $i$, $\mathbf{\Psi}_{i}$ is independent of $\epsilon_i$ in the primary model \eqref{model:blackbox}. 
Notice, the intermediary outcome models inform our primary outcome model
via $F_l$ for $l\in \{1,2, \cdots, L\}$ that in turn determine the candidate set of explanatory variables in \eqref{second:stage:Lasso} for a downstream modeling of the clinical outcome through the selected set $E$.
Completing the model specification in a Bayesian framework, we impose a rich family of scale-mixture Gaussian priors $\pi(\cdot)$
 on our selection-aware parameters $\Beta_{\bar{E}}$ \citep{park2008bayesian} :
\begin{equation}
\label{prior:beta}
\Beta_{\bar{E}}\ \lvert  \ \eta^2_1, \eta^2_2, \cdots, \eta^2_{q} \sim N(0, \sigma^2\text{diag}(\eta^2_1, \eta^2_2, \cdots, \eta^2_q)); 
\end{equation}
and $\eta^2_j \sim \text{Exp}(2^{-1}\lambda^2)$ for  $j\in \bar{E}$.
This construction admits a flexible class of models based on the following aspects of our modeling assumptions.
\begin{remark}
\label{rem:1}
The mapping, $\mathcal{E}$ is allowed to be a black box as long as it depends on the data only through $E$ from our regularized selection algorithms. 
Importantly, the interplay between preexisting knowledge and the selected set of variables $E$ need not be explicitly specified for selection-aware inference;
all we need for our inferential methods post selection is the value of this mapping, $\bar{E}$.
\end{remark}
\begin{remark}
\label{rem:2}
Second, the validity of our inferential approach is not tied to \eqref{model:intermediary}-- the selected linear model between the explanatory and intermediary variables. 
In this context, our methods rely on the independence between the model errors $\epsilon_i$ and $\mathbf{\Psi}_i$ for each data sample;
the proof for Proposition \ref{cond:lik} in the next Section justifies this observation.
We specify a linear model for the intermediary outcomes only for the sake of simple exposition in the remaining paper. 
\end{remark}

\vspace{-5mm}

\section{Selection-aware posterior inference}
\label{methods}

In the current section, we introduce a selection-aware posterior and then provide our inferential theory in Theorems \ref{correction:factor:cond} and \ref{reparam:map} to enable tractable updates when gradient-based MCMC sampling is used for estimating uncertainties from a working version of this posterior.

\subsection{Selection-aware posterior}
\label{sel:post}
Recall, the model in \eqref{model:blackbox} is dependent on the sets of selected variables $F_l$ for $l\in \{1,2,\cdots, L\}$ and $E$.
We call the respective random variables $\widehat{E}(\mathbf{Y}, \mathbf{R}) $ and $\widehat{F}_l(\mathbf{I}_l)$, highlighting their dependence on the outcome variable $\mathbf{Y}$, the intermediary variable $\mathbf{I}_l$ and the randomization variable $\mathbf{R}$. 
That is, we observe the following realizations from the model selection pipeline in Section \ref{motivation:model}:
\begin{equation}
\label{selections:data}
\widehat{E}(\mathbf{Y}, \mathbf{R})= E  \text{ and } \widehat{F}_l(\mathbf{I}_l)= F_l \text{ for } l=1,2,\cdots, L.
\end{equation}
Explicitly accounting for the selection-aware nature of our modeling framework, a likelihood conditioned upon observing \eqref{selections:data} discards the information from our samples utilized for model selection. We provide in Proposition \ref{cond:lik} the form of the conditional likelihood in terms of the parameters within the primary outcome model \eqref{model:blackbox}. 

\begin{Proposition}
\label{cond:lik}
Let $\bar{E}$ be defined according to \eqref{mapping}. Let $\widehat{\Beta}_{\bar{E}} =  (\mathbf{G}_{\bar{E}}^T \mathbf{G}_{\bar{E}})^{-1} \mathbf{G}_{\bar{E}} \boldsymbol{\mathrm{y}}$ be the least squares estimate  after regressing $\boldsymbol{\mathrm{y}}$ against $\mathbf{G}_{\bar{E}}$, with covariance matrix $\mathbf{\Sigma}_{\bar{E}}$. 
Then, under the modeling assumptions \eqref{model:blackbox} and \eqref{model:intermediary}, the likelihood obtained by conditioning the law of $\widehat{\Beta}_{\bar{E}}$ upon the observed event of selection in \eqref{selections:data} agrees with:
\begin{equation*}
\begin{aligned}
& \left\{\mathbb{P}[\widehat{E}(\mathbf{Y}, \mathbf{R}) = E  \;\lvert \;\Beta_{\bar{E}}] \right\}^{-1} \cdot \rho(\widehat{\Beta}_{\bar{E}}; \Beta_{\bar{E}}, \mathbf{\Sigma}_{\bar{E}})
\end{aligned}
\end{equation*}
up to a proportionality constant in $\Beta_{\bar{E}}$.
\end{Proposition}

Underscored in Section \ref{schematic:related}, the selection-aware posterior that appends the conditional likelihood in Proposition \ref{cond:lik} with a prior for $\Beta_{\bar{E}}$ is obtained by applying a multiplicative correction term to the usual posterior.
Basing inference on the selection-aware posterior is a formidable challenge, because the value of the posterior, due to conditioning, involves the probability of selection
\begin{equation}
\label{prob}
\mathbb{P}[\widehat{E}(\mathbf{Y}, \mathbf{R})= E  \;\lvert \;\Beta^{(d)}_{\bar{E}}]
\end{equation}
which must be computed in each new draw $\Beta^{(d)}_{\bar{E}}$ sampled from the posterior. 
The event of selection in \eqref{prob} as we characterize next is associated with an intricate geometry and the probability for the event involves integrating out several variables to render an exact value.
Circumventing geometric and analytic impediments to selection-aware inference, our solution in the following development casts the core step as an easy-to-solve, low dimensional, convex optimization problem  [\text{\normalfont \textbf{OP}}].

\subsection{A simplified conditional likelihood}
\label{cond:lik:sec}

Before stating our solution, Proposition \ref{polytope} characterizes the event of selection $\{(\boldsymbol{\mathrm{y}}, \boldsymbol{\mathrm{r}}): \widehat{E}(\boldsymbol{\mathrm{y}}, \boldsymbol{\mathrm{r}}) = E\}$ as a union of polyhedral regions determined by 
\begin{equation}
\label{U,V,W,t}
\left\{\mathbf{U}_{\mathbf{s}_E}, \mathbf{V}_{\mathbf{s}_E}, \mathbf{W}_{\mathbf{s}_E}, \mathbf{t}_{\mathbf{s}_E}: \mathbf{s}_E= \{-1,1\}^{|E|}\right\}.
\end{equation}
Detailed expressions for \eqref{U,V,W,t} are included in the Supplementary material.

\begin{Proposition}
\label{polytope}
After solving \eqref{second:stage:Lasso}, the selection event $\{(\boldsymbol{\mathrm{y}}, \boldsymbol{\mathrm{r}}): \widehat{E}(\boldsymbol{\mathrm{y}}, \boldsymbol{\mathrm{r}}) = E\}$  is equivalent to 
$$\underset{{\mathbf{s}_E= \{-1,1\}^{|E|}}}{\bigcup}\left\{\mathbf{U}_{\mathbf{s}_E}\widehat{\Beta}_{\bar{E}}+ \mathbf{V}_{\mathbf{s}_E}\begin{pmatrix}\boldsymbol{\mathrm{r}}^T_E &  \boldsymbol{\mathrm{r}}^T_{E^c}\end{pmatrix}^T + \mathbf{W}_{\mathbf{s}_E}\begin{pmatrix}(\widehat\Beta^{\perp}_E)^T &  (\widehat\Beta^{\perp}_{E^c})^T \end{pmatrix}^T> \mathbf{t}_{\mathbf{s}_E}\right\},$$
where $\widehat\Beta^{\perp}= \mathbf{G}_{\bar{F}}^T \boldsymbol{\mathrm{y}}- \mathbf{G}_{\bar{F}}^T \mathbf{G}_{\bar{E}} \widehat{\Beta}_{\bar{E}}$ and $E^c= \bar{F}\setminus E$.
\end{Proposition}

 By recognizing next a careful conditioning event, we reduce the seemingly complicated probability of selection, equivalent to the probability of a union of polyhedral regions, to that of an orthant based on very simple sign restrictions on our data variables.
 This results in a considerably simpler conditional likelihood function which we formalize in  
 Theorem \ref{correction:factor:cond}. 
Establishing some more notations, the stationary equation at the solution of \eqref{second:stage:Lasso} is given by:
\begin{equation}
\label{KKT}
\begin{aligned}
&\begin{pmatrix} \boldsymbol{\mathrm{r}}_E^T & \boldsymbol{\mathrm{r}}^T_{E^c} \end{pmatrix}^T + \begin{pmatrix} (\mathbf{G}_E^T \boldsymbol{\mathrm{y}})^T & (\mathbf{G}_{E^c}^T \boldsymbol{\mathrm{y}})^T  \end{pmatrix}^T  \\
&\;\;\;\;\;\;\;\;\;\;= \begin{bmatrix} \mathbf{G}_E^T \mathbf{G}_E + \epsilon \cdot \mathbf{I} & \mathbf{G}_E^T \mathbf{G}_{E^c}\end{bmatrix}^T \widehat{\Beta}_{E}^{\;\text{LASSO}}+\begin{pmatrix}  (\mathbf{\Lambda}_E \mathbf{s}_E)^T & (\mathbf{\Lambda}_{E^c} \mathbf{z} )^T\end{pmatrix}^T,
\end{aligned}
\end{equation}
where 
the active solution, $\widehat{\Beta}_{E}^{\;\text{LASSO}}$, and the inactive part of the subgradient, $\mathbf{z}$, satisfy the constraints
\begin{equation*}
\text{sign}( \widehat{\Beta}_{E}^{\;\text{LASSO}}) = \mathbf{s}_E\; ; \; \|\mathbf{z}\|_{\infty}< 1;
\end{equation*}
$\mathbf{s}_E$ is the vector of signs for the active (nonzero) LASSO solution.
We use $\widehat{\mathbf{B}}^{\text{LASSO}}$, $\widehat{\mathbf{Z}}$, and $\widehat{\mathbf{B}}^{\perp}$ to represent the random variables that assume the realizations $\widehat{\Beta}_{E}^{\;\text{LASSO}}$, $\mathbf{z}$, and $\widehat\Beta^{\perp}$ respectively upon solving \eqref{second:stage:Lasso}.
We defer the explicit forms for the matrices $\mathbf{P}$, $\mathbf{Q}$, $\mathbf{o}$, $\mathbf{K}$, $\mathbf{l}$ and $\boldsymbol{\Theta}_{\bar{E}}$ in the next result to the Supplementary material.

\begin{theorem}
\label{correction:factor:cond}
Consider the modeling assumptions in \eqref{model:blackbox} and \eqref{model:intermediary}.
Define 
\begin{equation*}
I(\Beta_{\bar{E}})= \int_{\text{sign}(w)= \mathbf{s}_E} \rho(b; \mathbf{K}\Beta_{\bar{E}}+\mathbf{l}, \boldsymbol{\Theta}_{\bar{E}})\cdot \rho(w; \mathbf{P}b+\mathbf{o}, \eta^{-2}\mathbf{Q}^T \mathbf{Q}) dw db
\end{equation*}
Then, conditional upon 
$$\left\{\widehat{E}=E, \ \text{sign}(\widehat{\mathbf{B}}^{\text{\normalfont LASSO}}) = \mathbf{s}_E,\ \widehat{\mathbf{Z}}= \mathbf{z},\ \widehat{\mathbf{B}}^{\perp}=  \widehat\Beta^{\perp}\right\},$$
 the likelihood of
the least squares estimate $\widehat\Beta_{\bar{E}}$ in Proposition \ref{cond:lik}
is proportional to 
\begin{equation}
\label{cond:lik:simple}
\begin{aligned}
&\left(I(\Beta_{\bar{E}})\right)^{-1}\cdot \rho(\widehat{\Beta}_{\bar{E}}; \mathbf{K}\Beta_{\bar{E}}+\mathbf{l}, \boldsymbol{\Theta}_{\bar{E}}).
\end{aligned}
\end{equation}
\end{theorem}

To further avoid carrying out the integration to calculate $I(\Beta_{\bar{E}})$ that lacks a value in closed form, we apply the Laplace technique \citep{tierney1986accurate, kass1995bayes} for approximating:
\begin{equation*}
\begin{aligned}
\log I(\Beta_{\bar{E}}) &\approx - \underset{{b, w}}{\inf} \; \Big\{\;\frac{1}{2} (b-\mathbf{K}\Beta_{\bar{E}}-\mathbf{l})^T \boldsymbol{\Theta}_{\bar{E}}^{-1}(b-\mathbf{K}\Beta_{\bar{E}}-\mathbf{l})\\
&\;\;\;\;\;\;\;\;\;\;\;+ \frac{1}{2\eta^2}(w- \mathbf{P}b-\mathbf{o})^T \mathbf{Q}^T \mathbf{Q} (w-\mathbf{P}b-\mathbf{o})  + \text{Barr}_{\mathbf{s}_E}(w)\Big\} +C ;
\end{aligned}
\end{equation*}
$C$ is a constant, and $w_j$ and $s_{j;E}$ are the values of the $j$-th coordinate of the respective vectors and $\text{Barr}_{\mathbf{s}_E}(w)= \sum_{j=1}^{|E|} \text{Barr}_{j;\mathbf{s}_{j,E}}(w_j)$ is a barrier penalty that encodes the sign constraints on each coordinate of $w$ through a smooth function $ \text{Barr}_{j;\mathbf{s}_{j;E}}(w_j) = \log (1+ \delta/\mathbf{s}_{j;E} w_j)$.  
In conjunction with our prior, the approximate value of $I(\Beta_{\bar{E}})$ plugged into the likelihood in \eqref{correction:factor:cond} leads us to a working version for the (log-) posterior:
\begin{equation}
\label{approx:posterior}
\begin{aligned}
&[\text{\textbf{SaP}}]: \ \log\rho(\widehat{\Beta}_{\bar{E}}; \mathbf{K}\Beta_{\bar{E}}+\mathbf{l}, \boldsymbol{\Theta}_{\bar{E}}) +\underset{{b, w}}{\inf} \; \Big\{\frac{1}{2} (b-\mathbf{K}\Beta_{\bar{E}}-\mathbf{l})^T \boldsymbol{\Theta}_{\bar{E}}^{-1}(b-\mathbf{K}\Beta_{\bar{E}}-\mathbf{l})\\
&\;\;\;\;\;\;\;\;\;\;\;\;\;\;\;\;+ \frac{1}{2\eta^2}(w- \mathbf{P}b-\mathbf{o})^T \mathbf{Q}^T \mathbf{Q} (w-\mathbf{P}b-\mathbf{o})  + \text{Barr}_{\mathbf{s}_E}(w)\Big\} + \log \pi(\Beta_{\bar{E}})
\end{aligned}
\end{equation}
after ignoring constants.

\subsection{Reparameterization mapping}
\label{log:posterior:exp}

We develop a reparameterization mapping in the current section to enable tractable updates from  \eqref{approx:posterior} through a convex optimization problem in $|E|$ dimensions when gradient-based MCMC sampling is deployed for inference.
With no additional cost, we can easily reconstruct using the same mapping our original targets, the parameters in the radiogenomic clinical model.

Consider the optimization:
\begin{equation}
 [\text{\normalfont \textbf{OP}}]:  \mathbf{w}^*(\Zeta_{\bar{E}}) =\underset{w}{\text{argmin}} \;\; \frac{1}{2\eta^2}(w-\mathbf{P}\Zeta_{\bar{E}} - \mathbf{o})^T \mathbf{Q}^T \mathbf{Q} (w- \mathbf{P}\Zeta_{\bar{E}} - \mathbf{o}) + \text{Barr}_{\mathbf{s}_E}(w),
\label{opt:update}
\end{equation}
denoting the optimal value by $V^*(\Zeta_{\bar{E}})$.
Based on the solution of \eqref{opt:update}, fix 
\begin{equation*}
 \Psi(\Zeta_{\bar{E}}) = \left(\mathbf{I} + \eta^{-2}\boldsymbol{\Theta}_{\bar{E}} \mathbf{P}^T \mathbf{Q}^T \mathbf{Q} \mathbf{P}  \right)\Zeta_{\bar{E}} + \eta^{-2} \boldsymbol{\Theta}_{\bar{E}} \mathbf{P}^T \mathbf{Q}^T  \mathbf{Q}(\mathbf{o}- \mathbf{w}^*(\Zeta_{\bar{E}})).
\end{equation*}
Then, we define the reparameterization $\Beta_{\bar{E}} \to \Zeta_{\bar{E}}$ through $\mathbf{K}$ and $\mathbf{l}$ and the mapping $\Psi(\cdot)$ as follows:
\begin{equation}
\label{reparam}
 [\text{\textbf{RP}}]: \mathbf{K}\Beta_{\bar{E}} + \mathbf{l}= \Psi(\Zeta_{\bar{E}}).
\end{equation}
Applying \eqref{reparam} to the working version for the (log-) posterior in \eqref{approx:posterior}, the next Theorem provides the value of a transformed analog for the working posterior $\widetilde{\pi}(\Zeta_{\bar{E}} \lvert \widehat{\Beta}_{\bar{E}})$ and the corresponding gradient in terms of the variables $\Zeta_{\bar{E}}$.

\begin{theorem}
\label{reparam:map}
Consider the reparameterization mapping in \eqref{reparam}. Fix 
\begin{equation*}
\begin{aligned}
\boldsymbol{\mathcal{J}}(\Zeta_{\bar{E}}) &=   \mathbf{K}^{-1}\left(\mathbf{I} + \eta^{-2}\boldsymbol{\Theta}_{\bar{E}} \mathbf{P}^T \mathbf{Q}^T \mathbf{Q} \mathbf{P} \right)-\eta^{-4}\mathbf{K}^{-1}\boldsymbol{\Theta}_{\bar{E}} \mathbf{P}^T \mathbf{Q}^T  \mathbf{Q}\mathbf{N}^{-1}\mathbf{Q}^T \mathbf{Q} \mathbf{P}
\end{aligned}
\end{equation*}
and 
$$\boldsymbol{\mathcal{M}}=\text{vec}(\boldsymbol{\mathcal{M}}_1, \cdots, \boldsymbol{\mathcal{M}}_{q}),$$
where $\mathbf{N} = \eta^{-2} \mathbf{Q}^T \mathbf{Q}  + \grad^2 \text{\normalfont Barr}_{\mathbf{s}_E}(\mathbf{w}^*(\Zeta_{\bar{E}}))$ and $\boldsymbol{\mathcal{M}}_j$ takes the value
\begin{equation*}
\begin{aligned}
&\text{\normalfont{Trace}}\Big( \frac{1}{\eta^{4}}\boldsymbol{\mathcal{J}}^{-1}(\Zeta_{\bar{E}})  \mathbf{K}^{-1}\boldsymbol\Theta_{\bar{E}} \mathbf{P}^T \mathbf{Q}^T \mathbf{Q} \mathbf{N}^{-1}\Big(\text{\normalfont{diag}}(\grad^3 \text{\normalfont  Barr}_{1;\mathbf{s}_{E;1}}(w_1^*(\Zeta_{\bar{E}})), \cdots, \\
&\;\;\;\;\;\;\;\;\;\;\;\;\;\;\;\;\;\;\;\;\;\;\;\;\;\;\;\;\;\;\;\;\;\;\;\grad^3  \text{\normalfont  Barr}_{|E|;\mathbf{s}_{E;|E|}} (w_E^*(\Zeta_{\bar{E}})) \text{\normalfont{diag}}(\eta^{-2}[ \mathbf{N}^{-1} \mathbf{Q}^T \mathbf{Q} \mathbf{P}]_{j})\Big) \mathbf{N}^{-1} \mathbf{Q}^T \mathbf{Q} \mathbf{P}  \Big)
\end{aligned}
\end{equation*}
 for $j\in \{1,\cdots, \bar{E}\}$. Then, we have the following.\\
$\mathrm{(i)}$ The value of $\log\widetilde{\pi}(\Zeta_{\bar{E}} \lvert \widehat{\Beta}_{\bar{E}})$, up to an additive constant, is equal to
\begin{equation*}
\begin{aligned}
&\log |\text{{\normalfont det}}(\boldsymbol{\mathcal{J}}(\Zeta_{\bar{E}}))| + ( \widehat{\Beta}_{\bar{E}}-\Zeta_{\bar{E}}) ^T \boldsymbol{\Theta}^{-1}_{\bar{E}} \Psi(\Zeta_{\bar{E}}) + \frac{1}{2}\Zeta_{\bar{E}}^T  \boldsymbol{\Theta}^{-1}_{\bar{E}}  \Zeta_{\bar{E}} + V^*(\Zeta_{\bar{E}}) \\
&\;\;\;\;\;\;\;\;\;\;\;\;\;\;\;\;\;\;\;\;\;\;\;\;\;\;\;\;\;\;\;\;\;\;\;\;\;\;\;\;\;\;\;\;\;\;\;\;\;\;\;\;\;\;\;\;\;\;\;\;\;\;\;\;\;\;\;\;\;\;\;\;  + \log \pi(\mathbf{K}^{-1}\Psi(\Zeta_{\bar{E}})-\mathbf{K}^{-1}\mathbf{l}).
\end{aligned}
\end{equation*}
$\mathrm{(ii)}$ The gradient for $\log\widetilde{\pi}(\Zeta_{\bar{E}} \lvert \widehat{\Beta}_{\bar{E}})$ is equal to
\begin{equation*}
\begin{aligned}
&  {(\boldsymbol{\mathcal{J}}(\Zeta_{\bar{E}}))^T\left(\grad \log \pi(\mathbf{K}^{-1}\Psi(\Zeta_{\bar{E}})-\mathbf{K}^{-1}\mathbf{l}) +  \mathbf{K}^T \boldsymbol{\Theta}^{-1}_{\bar{E}} (  \widehat{\Beta}_{\bar{E}} -  \Zeta_{\bar{E}})\right) + \boldsymbol{\mathcal{M}}(\Zeta_{\bar{E}}).}
\end{aligned}
\end{equation*}
\end{theorem}

With a sample $\Zeta^{(d)}_{\bar{E}}$ from $\widetilde{\pi}(\cdot \lvert \widehat{\Beta}_{\bar{E}})$, we obtain the corresponding draw for our target parameter $ \Beta^{(d)}_{\bar{E}}$ using the relation \eqref{reparam}. 
Notice, a reconstruction of our original target parameters does not involve an additional cost, because the primary cost is incurred in solving for $\mathbf{w}^*(\Zeta^{(d)}_{\bar{E}})$ which we compute to generate samples from the transformed posterior. 
Viewing this from the perspective of inferential efficiency, the optimization at every update of the transformed posterior is only $|E|$ dimensional, orders smaller in magnitude than $p$, the size of the initial set of explanatory variables.

\section{Simulation analysis}
\label{simulation}
We turn our attention to a reconciliation between the selection-aware nature of our Bayesian models and the validity of our inferential estimates in the empirical
analyses below. 
We discuss our design of experiment, demonstrate the potential of our methods in striking a balance between the number of discoveries and the power of discovery and 
 illustrate how these inferential metrics successfully generalize to dimensions beyond the radiogenomic application under study.

\subsection{Simulation design}
Generating a sparse model with both weak and strong signals of varying amplitudes and random signs, we draw in each round of simulation an outcome from the primary model \eqref{model:blackbox}.
The signal vector $\Beta$ is generated from a mixture of centered Laplace distributions, the true underlying prior. That is, each coordinate for $\Beta$ is drawn as follows: 
\begin{equation}
\label{signal:generation}
\beta_j \sim \Pi_j(\cdot):=  \pi \cdot \text{Laplace}(0, 0.10) + (1-\pi) \cdot \text{Laplace}(0, s) \text{ for }  j \in \{1,2,\cdots, r\};
\end{equation}
$r= |\bar{F}|$.
Changing the scale of one of these Laplace distributions, $s$ and the mixing proportion, $\pi$ results in different signal regimes; in particular, the mixing proportion controls the sparsity levels of our signal vector.

We vary in our design the ratio between the number of our samples, $n$ and the number of regressors, $r$ before selecting the set $E$ that in turn determines the model \eqref{model:blackbox}. 
For the real data analysis, we note $r/n \approx 5$, $n=60$, $r=357$. 
In this case, the matrix of explanatory variables we use is based on the real values of genomic measurements that are significantly associated with the imaging (radiomic) outcomes.
We provide a comprehensive background for this data in the next section.
Beyond the real setting, we investigate the following sample sizes: $n=180,\ 360, \ 720$, to match the dimension ratio $r/n = 2, 1, 0.5$ respectively. 
To generate predictor measurements for sample sizes larger than $60$, we append synthetic design values to the real design matrix, in order to achieve the regression dimensions set as per the ratio $r/n$. Specifically, we draw $x_i \in \mathbb{R}^r, 1\leq i \leq n_1$ such that $x_i\sim N(0, \boldsymbol\Sigma(\rho))$, for $i=1,2,\cdots, n_1$ and $n_1$ is chosen so that $r/(n_1+n)= 2, 1, 0.5$ in the three case studies of interest; $\boldsymbol\Sigma(\rho)$ is an autocorrelation covariance matrix such that the $(i,j)$-th entry of $\boldsymbol\Sigma$ is equal to $\rho^{|i-j|}$ and $\rho=0.70$.

Our strategy to reflect a realistic data generation process through simulations is aligned along the following principles. First, we note that the variability in the outcome variable is explained by multiple markers, consistent with our expectation of a polygenic response variable. Second, the generative model we use incorporates a mix of weak and strong signals with varying amplitudes and random signs. This enables us to investigate the genuine ability of our inferential methods to adapt to the strength of the signals present in the data and to reconstruct efficiently the corresponding effect sizes. Third, admitting a range of regression dimensions, our simulations showcase (i) the necessity of adopting our methods even in moderate dimensions where a severe impact of selection bias is seen; (ii) the larger number of discoveries that our methods support, with reduced variance for the associated inferential estimates than the benchmark offered by splitting.

\subsection{Empirical analysis: consistent with radiogenomic case study}

\begin{figure}
  \centering
  \begin{tabular}{@{}c@{}}
    \includegraphics[height=155pt,width=1.02\linewidth]{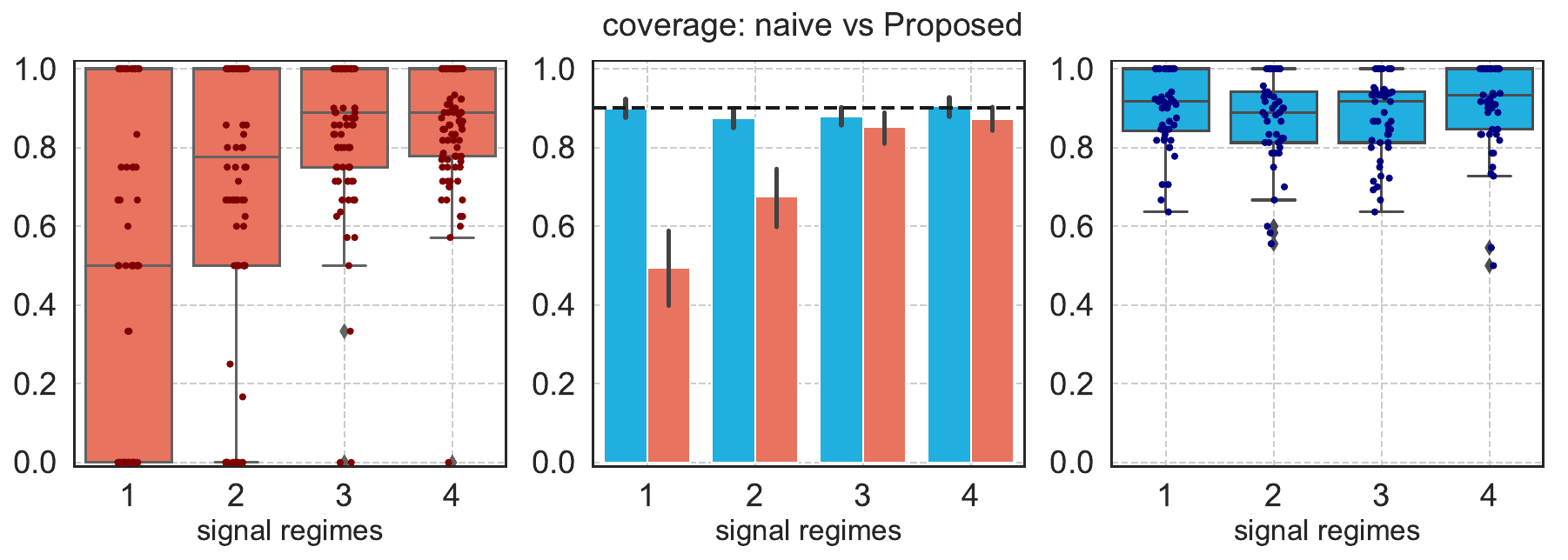} \\ [\abovecaptionskip]
    \small (I) 
      \end{tabular}
  
  \vspace{\floatsep}
  
  \begin{tabular}{@{}c@{}}
    \includegraphics[height=172pt,width=1.02\linewidth]{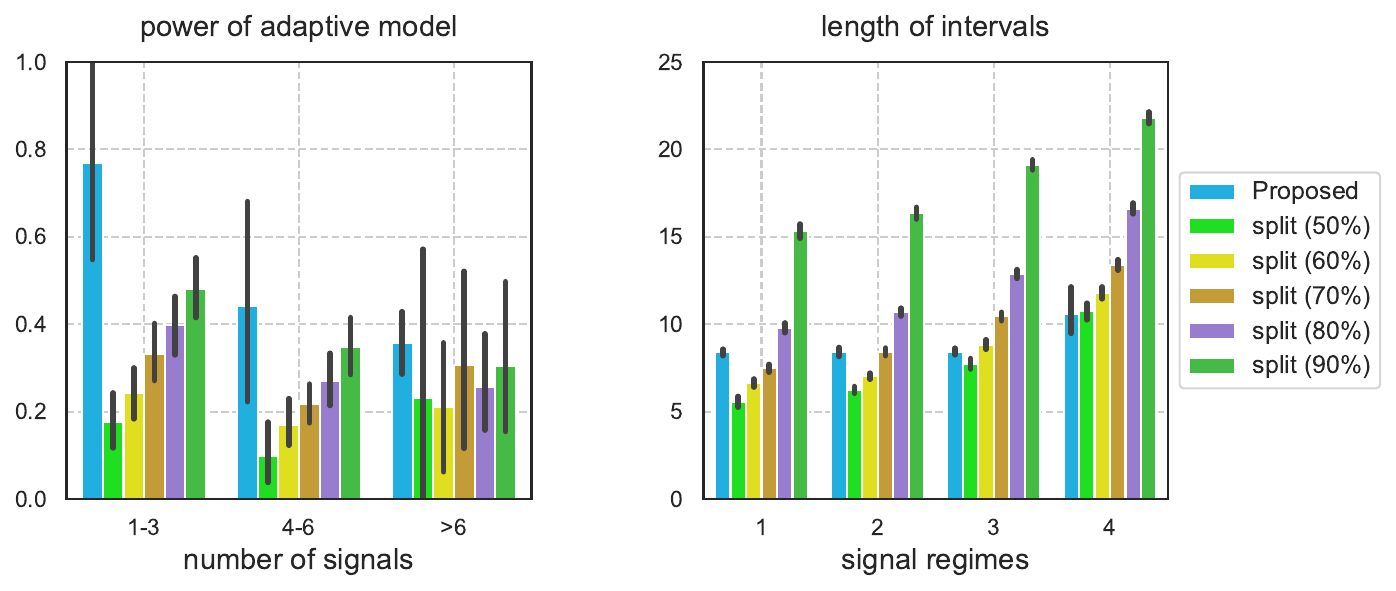} \\ [\abovecaptionskip]
    \small (II) 
  \end{tabular}
  \caption{(I): Invalidity of naive inference (in orange)-- the x-axis represents different signal regimes and the y-axis plots empirical coverage of $90\%$-intervals across the regimes. Left panel shows the distribution of naive estimates that ignore adaptive nature of model; central panel compares the averaged coverages across the regimes; right panel plots empirical distribution of the proposed intervals. The dotted line at $0.90$ is the nominal coverage. 
  (II): Tradeoff in model selection and inferential power-- left panel highlights the quality of our integrative model measured via the screening power of true signals; right panel shows the distribution of averaged lengths of interval estimates in the selected model.}
 \label{naive:proposed}
\end{figure}

We begin by exploring our methods with the real radiogenomic measurements, simulating $n=60$ samples to agree closely with the real data.
We set the randomization variation $\eta^2$ in \eqref{second:stage:Lasso} to be equal to the noise level in the outcome by plugging in an estimate of this value from the data. 
Setting $\pi= 0.95$ in the generative scheme \eqref{signal:generation} and varying the scale of the Laplace distribution $s$ to take one of the values in the set $\{0.20, 1, 2, 4\}$, we consider $4$ signal regimes-- numbered $1$-$4$ on the x-axis of Figure \ref{naive:proposed}. 
Using the expressions for the posterior and the gradient in Theorem \ref{reparam:map}, we update our estimates for $\Beta_{\bar{E}}$ from the working version of the selection-aware posterior \eqref{approx:posterior} and construct intervals for these parameters by setting $\bar{E}=E$, the final output of the regularized variable selection algorithms.

As noted in the introduction, a balance in the quality of the model in terms of support recovery and inferential power for the matched parameters is imminent when a finite amount of data must be allocated for deciding a model and inferring for the parameters in it. 
Illustrating the validity of our interval estimates post selection, we present detailed comparisons for this tradeoff of information between our approach and sample-splitting at different resolutions. 

Figure \ref{naive:proposed} anchors the motivation behind using a selection-aware posterior in panel (I). The distribution of the empirical coverages of naive intervals that ignore the selection-aware nature of integrative models have averaged coverage falling way short of the benchmark target of $90\%$. The interval estimates furnished by our methods support the validity and necessity of the inferential proposal in the paper. 
Panel (II) exemplifies a significantly better reconciliation between the recovery of signals from the integrative model and the subsequent inferential power for the matched parameters, when compared against splitting based on varying proportions of data reserved for selection. Specifically, the performance of the rather unconventional randomized query \eqref{second:stage:Lasso} in terms of model selection is evaluated using the number of true signals screened under different sparse scenarios and the follow up inferential power is depicted as the lengths of interval estimates averaged across simulations in these signal regimes. 

Observe, splitting where $90\%$ of the data is assigned for selecting signals is the best performer amongst all the split-based methods in terms of model-selection. However, this power is clearly dominated by the randomized scheme we adopt for modeling. In an assessment of inferential power, our methods accounting appropriately for the bias from model selection provide interval estimates which are less than half the length of the $90\%$ split-based intervals. 
 A take away from this illustration is the attractive alternative our methods offer in comparison to splitting across a range of resolution in terms of data allocation for the two core tasks in Panel (II). Evidently, our methods allow a distinctly unique yet more efficient tradeoff in the use of information for modeling and estimating uncertainties thereof.

\subsection{Inferential results: an illustration of our scope}
\begin{figure}
  \centering
  \begin{tabular}{@{}c@{}}
    \includegraphics[height=132pt,width=0.90\linewidth]{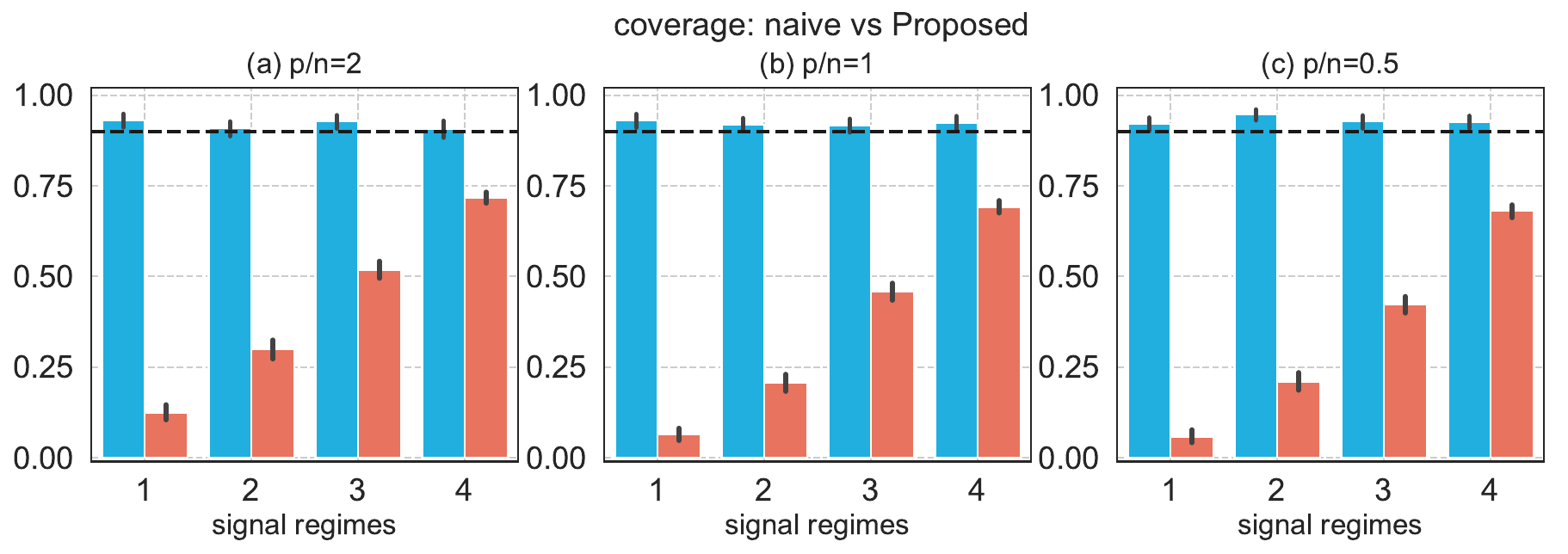} \\ [\abovecaptionskip]
    \small (I) 
      \end{tabular}
  
  \vspace{\floatsep}
  
  \begin{tabular}{@{}c@{}}
    \includegraphics[height=280pt,width=0.90\linewidth]{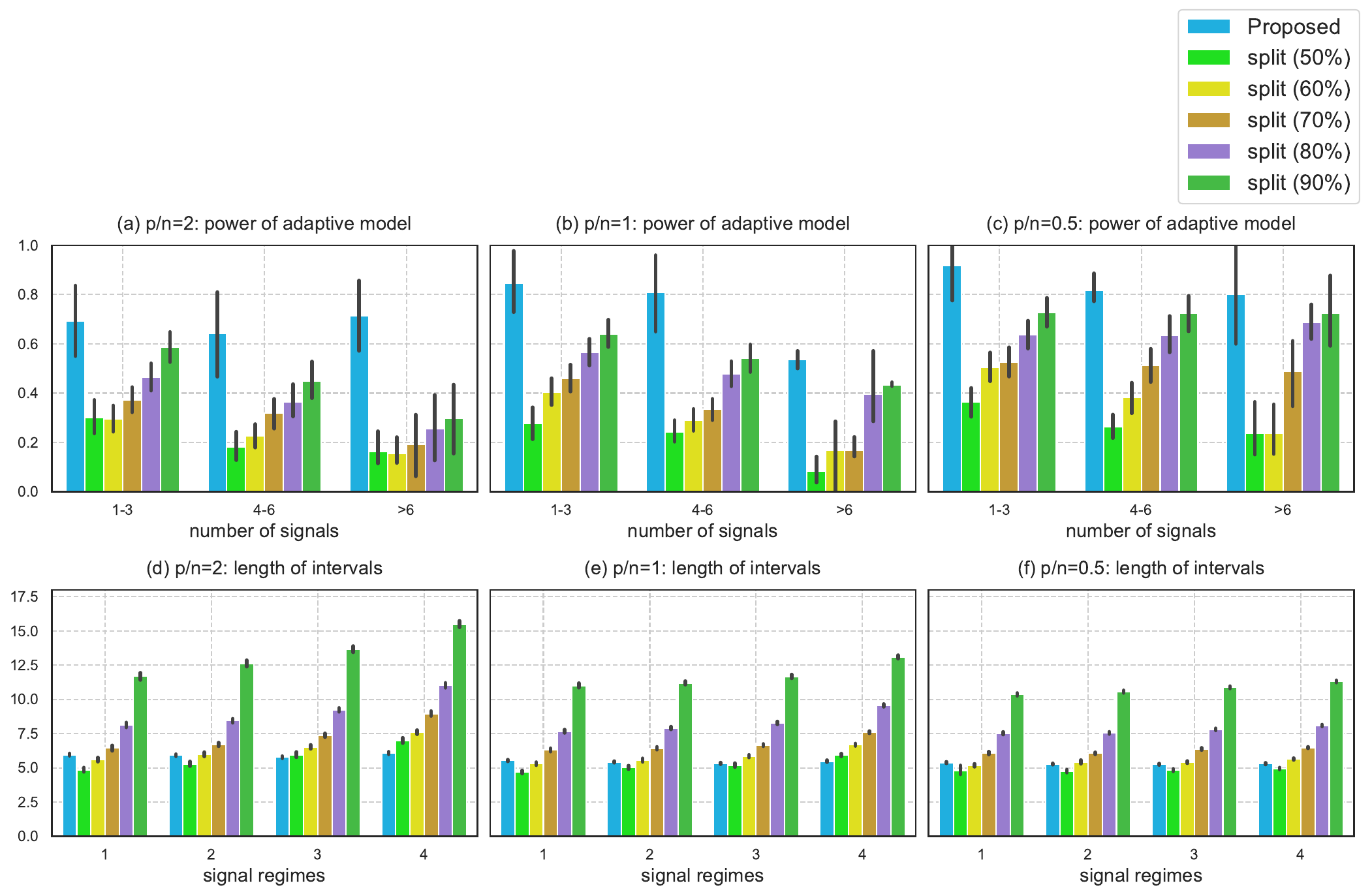} \\ [\abovecaptionskip]
    \small (II) 
  \end{tabular}
  \caption{(I): The x-axis represents different signal regimes; the y-axis plots empirical coverage of $90\%$-intervals with the dotted line at $0.90$ representing the target coverage. The panels depict the averaged coverages of naive and Proposed intervals for the dimensions $r/n=2, 1, 0.5$. 
  (II): Assessment of the screening power for the selected models in Panels (a), (b), (c), followed by the matching inferential power in Panels (d), (e), (f)— showcasing a tradeoff between selection and inference.}
 \label{naive:proposed:multiple}
\end{figure}

We next demonstrate how our methods generalize in their application to other data dimensions beyond our focused study.
The depiction in Figure \ref{naive:proposed:multiple}, through the averaged coverages of naive and Proposed intervals across the regression dimensions $r/n=2, 1, 0.5$, emphasizes the strong need to correct for selection bias .
For the signal regimes described in our simulation design, we see a severe shortfall of coverage for the naive interval estimates, ranging as low as $\sim 10\%$ and increasing to a level of only $\sim 70\%$ in the moderate SNR regimes. We remark here that the coverage of the naive intervals worsens in comparison to the case study in the previous discussion. This difference in the behavior of the naive intervals can be attributed to the synthetic predictor values we append to the real radiogenomic observations in the simulations in order to vary the size of regression in this design.

In panel (II) of the same Figure, we highlight 1) the sharpness of the selection-aware model in (a), (b), (c) in terms of the number of signals screened by the randomized strategy \eqref{second:stage:Lasso} and split-based schemes, 2) inferential power in (d), (e), (f) measured as the averaged lengths of the interval estimates produced by the Bayesian proposal in the paper when compared to splitting. Coherent with the findings in the preceding discussion, the proposed methods dominate all the split-based methods when assessed for the quality of the selected model; the percentage in the legend indicates the proportion of data samples reserved for model selection. 

In balancing the allocation of samples towards the two tasks of model selection and inference, splitting based on $90\%$ of the samples for selection expectedly produces the best model amongst the split-based strategies. Yet, this split-based method falls short of the randomized selection in terms of the quality of the primary model of interest. Comparing the averaged lengths of the interval estimates in the four signal regimes, we note that our selection-aware Bayesian inferential methods consistently yield intervals that are shorter by two-three times than split ($90\%$). On the other hand, choosing a split-based approach with $50\%$ of the samples devoted for model selection results in a relatively worse model for inference, leading to a lesser number of discoveries. Figure \ref{naive:proposed:multiple} summarizes the advantages our selection-aware techniques enjoy over the common practice of splitting the data into two parts. Applying conditional inference after randomizing corrects precisely for the bias from model selection and permits an optimal reuse of data samples at the same time.

\section{Radiogenomic analysis for LGG} 
\label{real:data}

In this section, we implement our selection-aware pipeline on the samples from the radiogenomic case study.
The imaging outcomes, also called radiomic phenotypes, are collectively harnessed in integrative models with the genomic measurements to assess associations with overall survival for the patients. 
 We briefly describe the data acquisition and pre-processing steps for both the imaging and genomic modalities with specifics largely deferred to Supplementary material \ref{section:3:Supplement}. 
 We then give biological insights into the radiogenomic findings from our integrative model, situating their relevance in the context of recent scientific literature.

    \subsection{Pathway scores and radiomic phenotypes}
    \label{subsec:genomic:mri}
	We obtain the genomic data from LinkedOmics \citep{vasaikar2017linkedomics}, a publicly available portal that includes multi-omics data from multiple cancer types in TCGA. The genomic data we acquire are normalized gene-level RNA sequencing data from the Illumina HiSeq system (high-throughput sequencing) with expression values in the $\log_2$ scale. Focusing on subjects with LGG, the gene expression data includes 516 samples and 20086 genes, which is narrowed down further to the intersecting samples with imaging phenotypes in an integrative study. 
	
	A set of genes broadly constitutes a gene pathway. In our case, we derive the pathway membership of genes from the Molecular Signature Database \citep{liberzon2011molecular}, a publicly available resource containing annotated gene-sets divided into multiple collections (groups of pathways). Particularly, we consider four collections namely Hallmark Pathways (50 pathways), KEGG Canonical Pathways (KEGG - 186 pathways), Cancer Gene Neighborhoods and Cancer Modules (C4 - 858 pathways), and Oncogenic Signatures (C6 - 189 pathways). In Supplementary material \ref{sec:pathway:scores}, we discuss the construction of pathway scores for these genomic measurements.
	
	For the imaging records, we obtain the pre-operative multi-institutional MRI scans of TCGA LGG collection available in TCIA \citep{clark2013cancer}. 
	For our analysis, we consider four types of MRI sequences which include (i) native (T1), (ii) post-contrast T1-weighted (T1Gd), (iii) T2-weighted (T2), and (iv) T2 fluid attenuated inversion recovery (FLAIR) volumes. Each of these sequences display different types of tissues with varying contrasts based on the tissue characteristics. From the whole brain MRI scans, the tumor regions can be identified using an automated segmentation method called GLISTRboost \citep{bakas2017advancing}.
	These segmentation labels additionally identify each voxel as one of the three tumor sub-regions namely, necrotic and non-enhancing tumor core (NC), the peritumoral edema (ED) and the enhancing tumor (ET). In Figure \ref{fig: mri}, we show an axial slice from the MRI scan of a LGG subject corresponding to all four imaging sequences as well as the segmented tumor sub-regions. 
	\begin{figure}[h]
		\centering
		\resizebox{\textwidth}{!}{%
		\begin{tabular}{cc}
		    \begin{tabular}{cc}
			\subfigure[T1]{\includegraphics[trim=1cm 1cm 1cm 2cm, clip, width=1.6in, height=1.6in]{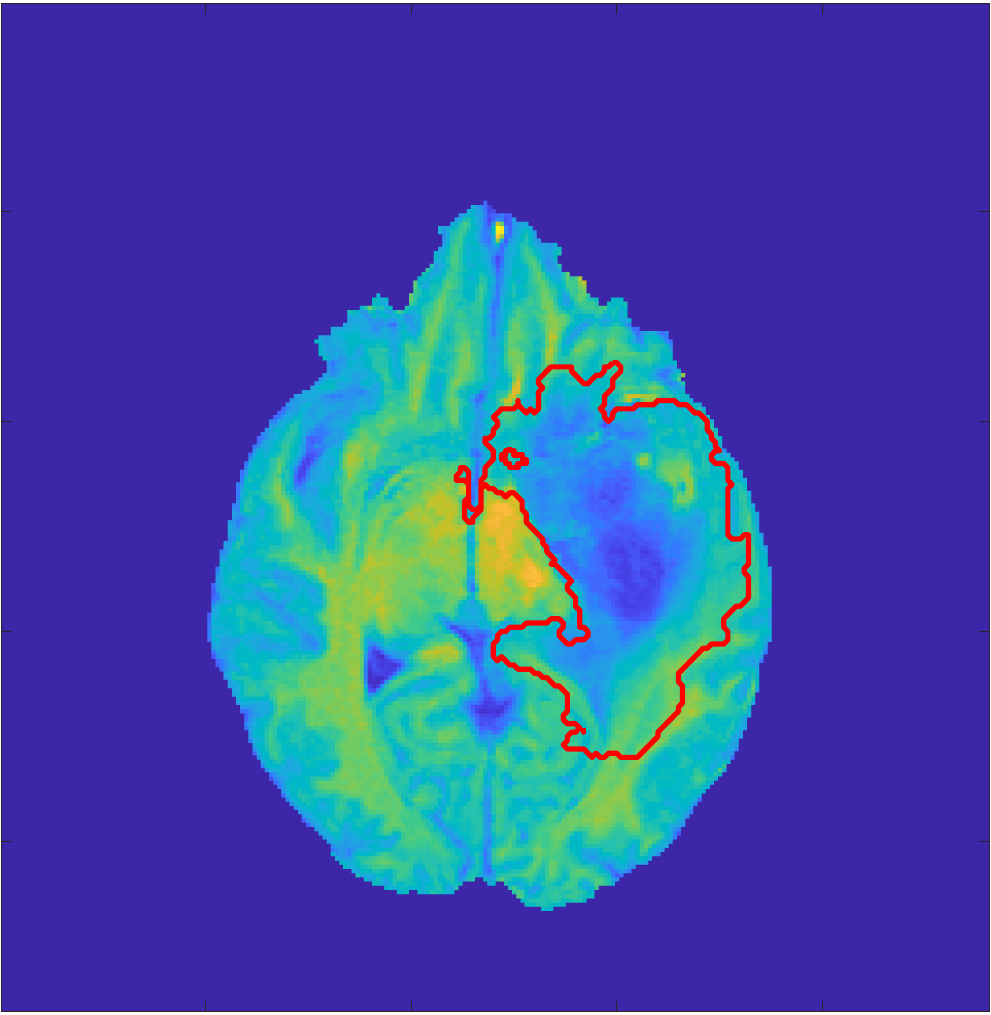}} & \subfigure[T1Gd]{\includegraphics[trim=1cm 1cm 1cm 2cm, clip, width=1.6in, height=1.6in]{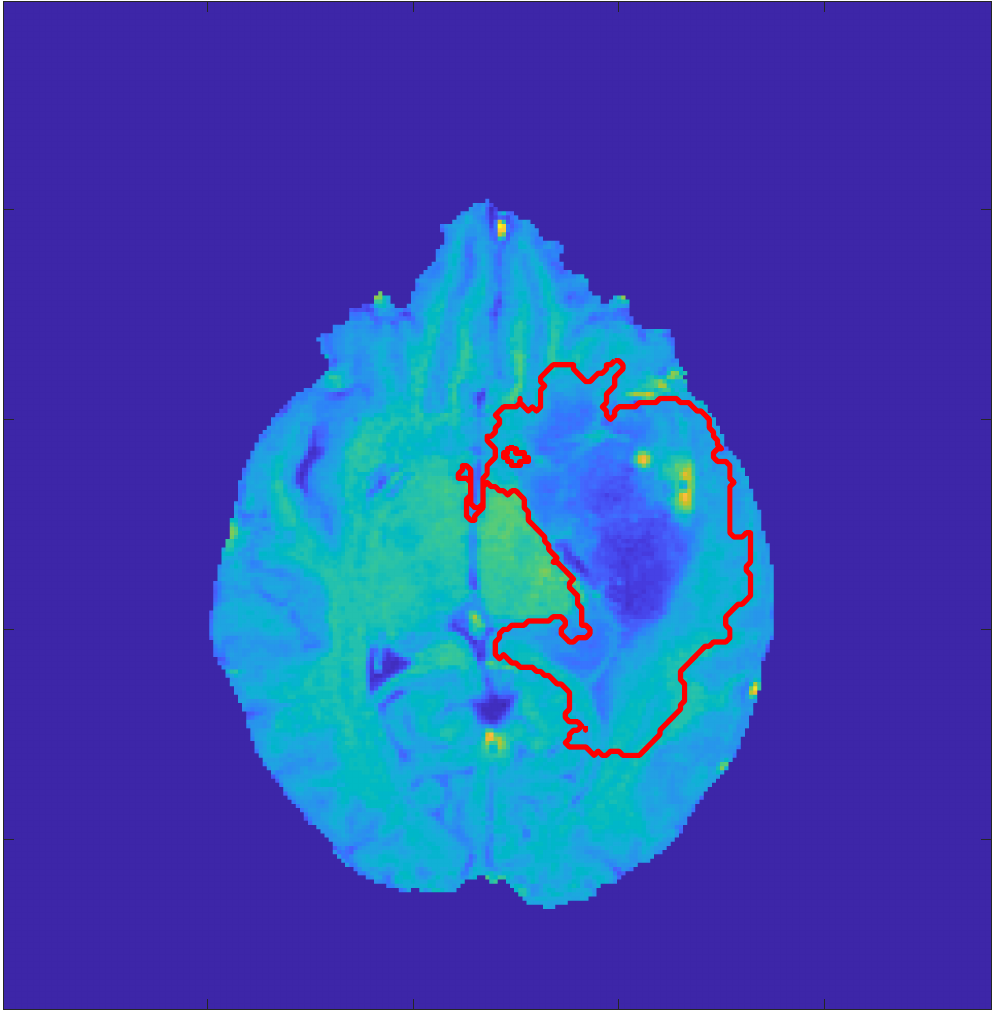}} \\
			\subfigure[T2]{\includegraphics[trim=1cm 1cm 1cm 2cm, clip, width=1.6in, height=1.6in]{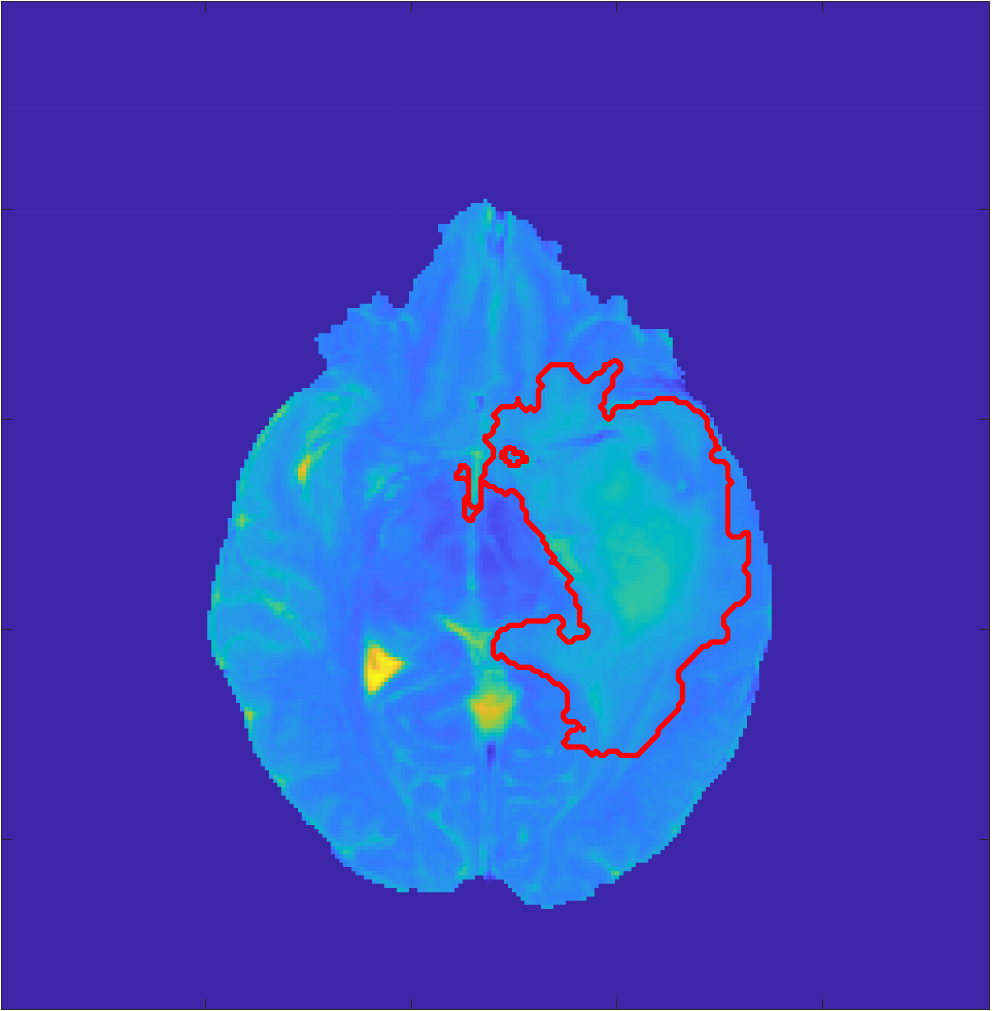}} & \subfigure[FLAIR]{\includegraphics[trim=1cm 1cm 1cm 2cm, clip, width=1.6in, height=1.6in]{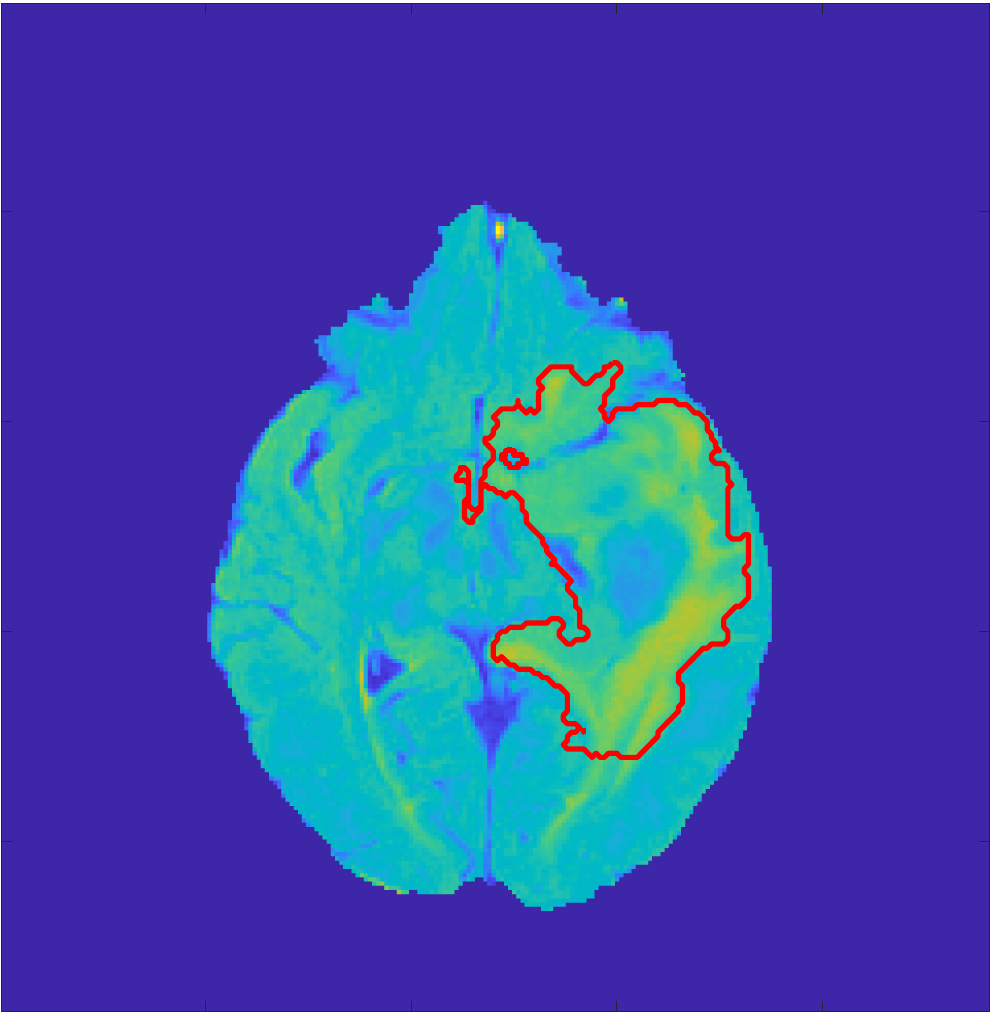}} \\
		\end{tabular} & \subfigure[Segmentation mask]{\includegraphics[trim=0.05cm 0.05cm 0cm 0cm, clip, width=1.6in, height=1.6in]{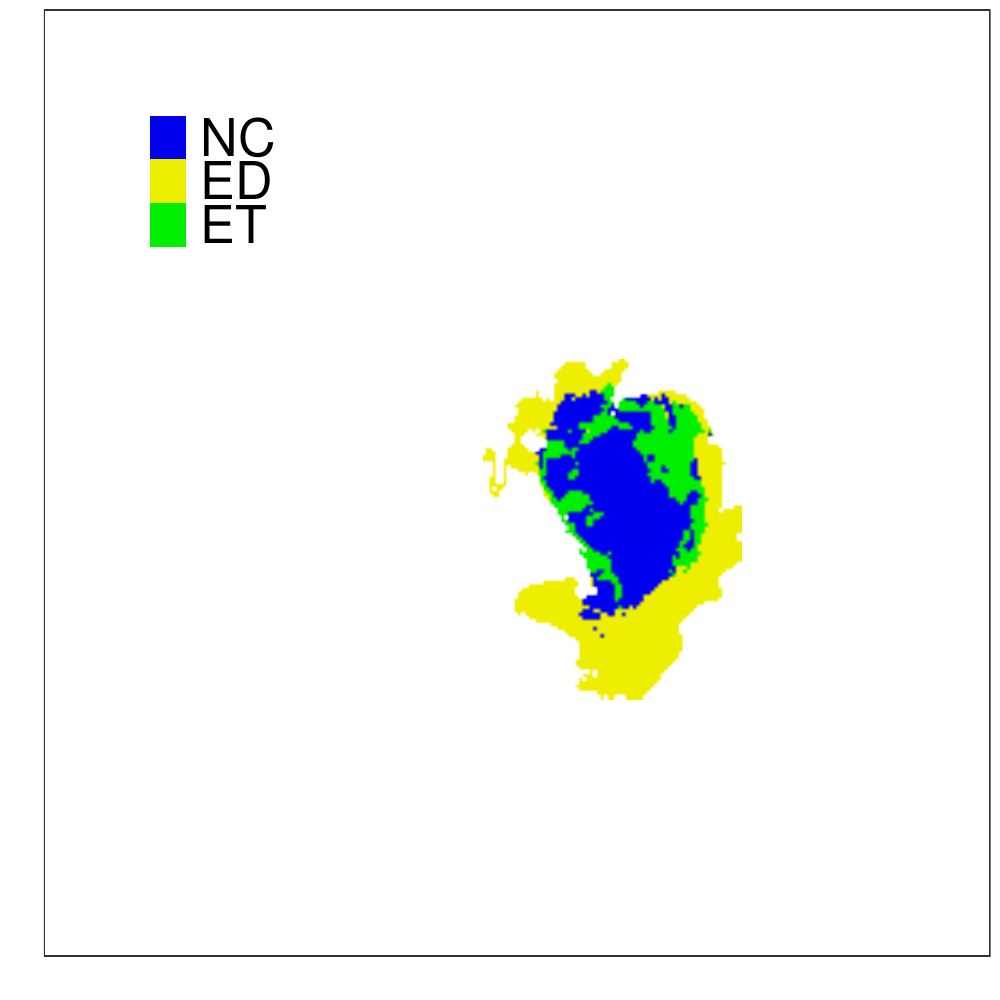}}
		\end{tabular}
		}
		\caption{An axial slice of a brain MRI from four modalities: T1, T1Gd, T2 and FLAIR for a LGG subject. The segmented tumor region is shown with an (red) overlaid boundary. Segmentation mask indicates the necrotic and non-enhancing tumor core (NC), the peritumoral edema (ED) and the enhancing tumor (ET) regions.}
		\label{fig: mri}
	\end{figure}

	The voxel intensity values so obtained are sensitive to the configuration of the MRI machine and are difficult to interpret. These values are neither comparable across different subjects, nor between study visits for the same subject. We address this issue by implementing a biologically motivated normalization technique called white-stripe normalization \citep{shinohara2014statistical}. Finally, we work with the intensity values corresponding to each of the 12 group of voxels (4 MRI sequences and 3 tumor sub-regions) separately. For each of the 12 groups of intensity values, we construct kernel density estimates for all the $61$ subjects. These densities give rise to the principal component scores for each of the 12 groups separately, whose details are included in Supplementary material \ref{subsec:pc_scores}.

    \subsection{Selection-aware pipeline with radiogenomic characteristics}\label{subsec:mining}
    The first step of the integrative selection pipeline identifies a set of promising pathways associated with the radiomic-based intermediary (imaging) outcomes. 
    To this end, we solve \eqref{first:stage:Lasso} with the $143$ principal component scores across the $12$ groups of tumor voxels as responses, regressed against
    $1289$ pathways from the four pathway collections (Hallmark, KEGG, C4, and C6). The output of this step is a set of $369$ gene pathways, each of which is associated with one or more of the radiomic phenotypes. 
    Of these 369 pathways selected, we note that the multiplicity of each pathway, defined as the number of LASSO queries which selects this potential predictor, ranges between $ 1-9$.

\begin{figure}[h]
		\centering
		\resizebox{\textwidth}{!}{
		\includegraphics[scale=1]{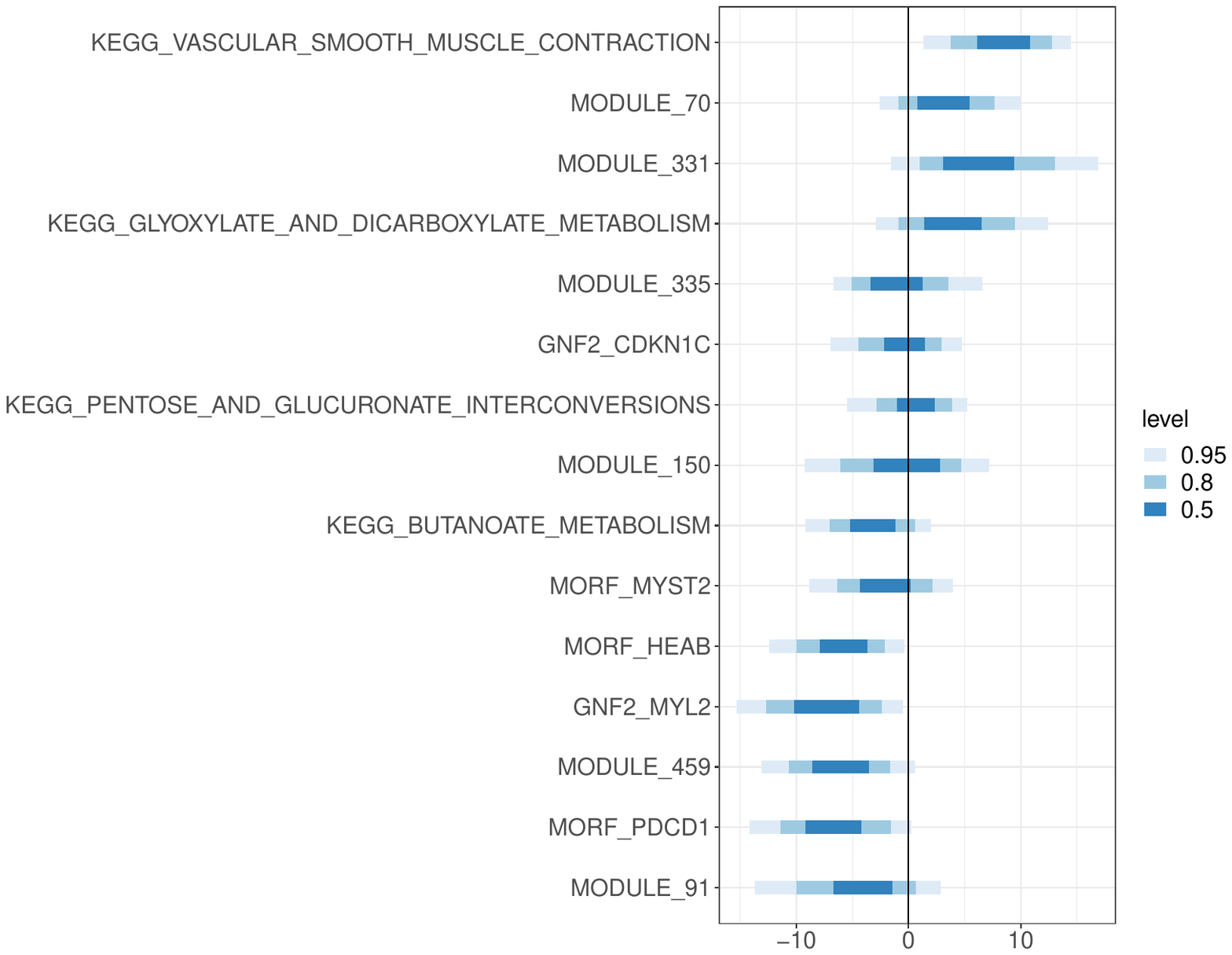}
		}
		\caption{Bounds of the credible intervals for the gene pathways associated with the overall survival and with the radiomic characteristics. }
		\label{fig:results}
	\end{figure}

    Using the log-transformed values of overall survival times as our clinical outcome, the second step solves a randomized version of LASSO \eqref{second:stage:Lasso} to partition the information within our data towards selecting a model and ascertaining strengths of these selected associations. 
    The penalty weights in the LASSO are set to be inversely proportional to the multiplicity of a pathway to reflect an importance weight for that feature in terms of its association with the imaging outcomes. 
    We select $15$ pathways from this step; these pathways are the imaging informed explanatory variables associated with survival. 
    Letting $\bar{E}=E$ in \eqref{model:blackbox} and using the prior in \eqref{prior:beta}, we use the optimization-based expressions for the selection-aware posterior to adjust for bias from the integrative selection pipeline. 
    Inference for the adaptively determined parameters $\Beta_E$ gives us effect size estimates for the pathways indexed by $E$. In Figure \ref{fig:results} we showcase the bounds for $50\%, 80\%$ and $95\%$ credible intervals based on the MCMC samples for the $15$ selected pathways.
    
\subsection{Biological Interpretations}\label{subsec:biological}
We now focus on some of our findings, providing their biological implications and interpreting the same in the context of existing clinical knowledge in this domain. 

\begin{enumerate}[leftmargin=*]
    \item We see that the gene pathway, Vascular Smooth Muscle Contraction, from the KEGG collection has significant association with overall survival. Vascular smooth muscle cell (VSMC) is a highly specialized cell whose principal function is contraction. These cells shorten on contraction, consequently decreasing the diameter of a blood vessel to regulate the blood flow and pressure. Moreover, in a clinically relevant mouse model of glioma, it was found that the glioma cells disrupt the VSMCs as they populate the perivascular space of preexisting vessels, causing a focal breach in the blood brain barrier \citep{watkins2014disruption}. It has been demonstrated that endothelial specific growth factor such as, vascular endothelial growth factor (VEGF), can interact with non-endothelial cells and play a role in modulating the response of VSMCs \citep{ishida2001expression}. VEGF expression levels were associated with the presence of ringlike tumor contrast enhancement, which present phenotypically as variable contrast on T1Gd MRI scan and were jointly associated with progression-free survival in glioblastoma \citep{wang2016radiologic}. 
    
    \item The gene pathway denoted as MORF PDCD1 which includes the genes in the neighborhood of the gene PDCD1, is seen to have a significant association with the overall survival. Recent work \citep{rover2018pd}  indicates that PDCD1 promoter methylation is a prognostic factor in LGG with Isocitrate Dehydrogenase (IDH) mutations. It is known that high expression of PDCD1 on the immune cells infiltrating the LGG is a marker for immune evasion and associated with survival.
    
    \item Another significant association we notice in our analysis corresponds to the gene pathway GNF2 MYL2. This is a group of genes in the neighborhood of MYL2 (myosin light chain II). Previous studies \cite{beadle2008role} show that myosin II plays a significant role in glioma invasion in vivo, where it regulates the deformation of the nucleus as well as the membrane of glioma cells. This has been further validated  through mathematical modeling in recent literature \citep{lee2017role}. Notably the extent of immune/inflammatory activity is reflected through the edema region on MRI scan \citep{kleijn2011distinguishing}.
\end{enumerate}

Some of the other pathways associated with the overall survival include metabolic pathways from KEGG such as (a) pentose and glucuronate interconversions, (b) glyoxylate and dicarboxylate metabolism, and (c) butanoate metabolism. In light of the significant role that metabolic reprogramming plays in glioma pathogenesis \citep{strickland2017metabolic}, it is encouraging to see that a number of metabolic pathways are identified to be significantly associated with patient prognosis. From the pathways for cancer gene neighborhoods we see significant associations with the gene pathways such as neighborhoods of (a) MYST2, a histone acetyltransferase that plays crucial functions in transcription, DNA replication and repair, and (b) CDKN1C, which is known to regulate several of the hallmark properties of cancer \citep{kavanagh2011hallmarks}. A follow-up validation of these pathways will illuminate nuances in understanding the tumor etiology under study.

\section{Concluding remarks}
\label{conclusion}
We conclude by remarking that there is certainly room for future directions. 
Integrative models may be prohibitive if there are important genomic variable(s) with a strong impact on the clinical outcome, but are culled out in the regression with the imaging outcomes.
Models attempting to link genomic variables directly with clinical endpoints might, however, be less viable, especially if the number of explanatory variables is many times larger than the number of available samples and subsets of these variables share substantial correlations.
In these situations, eliminating the upstream regression with the imaging outcomes may lead to a loss of accuracy as well as power in terms of support recovery.
Some discussion around a direct modeling approach (without the use of intermediary outcomes) is provided for our radiogenomic study in the Supplementary material. 
Recently, work by \cite{panigrahi2020approximate} propose selection-aware methods for regression with the Group LASSO penalty.
A generalization of our methods in this paper for regression with other structured penalties, in the integrative domain and on radiogenomic studies, are left as promising directions for future work.

\section{Acknowledgements}
S.P. acknowledges support through NSF-DMS 1951980 and NSF-DMS 2113342. S.M., A.R and V.B. were supported through CCSG P30 CA046592, Precision Health Scholar award (to S.M), Institutional Research Grants from The University of Michigan, NCI R37CA214955-01A1, and a Research Scholar Grant from the American Cancer Society (RSG-16-005-01). V. B. was also supported by NIH grants R01-CA160736,  R21-CA220299, NSF grant 1463233, and start-up funds from the U-M Rogel Cancer Center and School of Public Health.

\bibliographystyle{biom}
\bibliography{references.bib}

\renewcommand{\thesubsection}{\Alph{subsection}}
\section{Supplementary material}

\subsection{Proofs of main results}
\label{section:1:Supplement}

\begin{proof} \ \ Proposition \ref{cond:lik}. \ \ 
Using the following notations 
$$\boldsymbol{\mathcal{I}}_i=\begin{pmatrix} I_{i, 1} &\cdots & I_{i, L} \end{pmatrix}, \ \  \Mu_i(\Alpha_{\bar{F}}) = \begin{pmatrix} \mathbf{G}_{i; F_1} \Alpha_{F_1} & \cdots & \mathbf{G}_{i; F_L} \Alpha_{F_L} \end{pmatrix},$$ 
and using the independence between the clinical outcome, the intermediary outcome and the randomization variable,
we observe that the selection-ignorant (unconditional) likelihood under
 \eqref{model:blackbox} and \eqref{model:intermediary} is proportional to
 \medskip
\begin{equation}
\label{uncond:lik}
\begin{aligned}
&\exp\left(-\frac{1}{2\sigma^2}(\boldsymbol{\mathrm{y}} -\mathbf{G}_{\bar{E}}\Beta_{\bar{E}})^T (\boldsymbol{\mathrm{y}} -\mathbf{G}_{\bar{E}}\Beta_{\bar{E}})\right) \cdot \exp\left(-\frac{1}{2\eta^2}\boldsymbol{\mathrm{r}}^T\boldsymbol{\mathrm{r}}\right)\\
&\;\;\;\;\;\;\;\;\;\;\;\;\;\;\;\;\;\;\;\;\;\;\;\;\;\; \times  \exp\left(-\frac{1}{2}\sum_{i=1}^{n}(\boldsymbol{\mathcal{I}}_i- \Mu_i(\Alpha_{\bar{F}}))^T \mathbf{\Sigma}^{-1}_I (\boldsymbol{\mathcal{I}}_i-\Mu_i(\Alpha_{\bar{F}}))\right).
\end{aligned}
\end{equation}
Fix the following sets:
$$\mathcal{F}_0:=\left\{\boldsymbol{\mathcal{I}}: \widehat{F}_l(\boldsymbol{\mathcal{I}}_l) = F_l \text{ for }  l\in \{1,\cdots, L\}\right\};$$
$$\mathcal{E}_0 := \left\{(\boldsymbol{\mathrm{y}}, \boldsymbol{\mathrm{r}}): \widehat{E}(\boldsymbol{\mathrm{y}}, \boldsymbol{\mathrm{r}}) = E \right\}.$$
Truncating the law for the random variables $\mathbf{Y}$, $\mathbf{I}$, $\mathbf{R}$ in \eqref{uncond:lik} to the event resulting in the selected sets \eqref{selections:data}
yields us a joint conditional law proportional to:
\medskip
\begin{equation*}
\begin{aligned}
&\exp\left(-\frac{1}{2\sigma^2}(\boldsymbol{\mathrm{y}} -\mathbf{G}_{\bar{E}}\Beta_{\bar{E}})^T (\boldsymbol{\mathrm{y}} -\mathbf{G}_{\bar{E}}\Beta_{\bar{E}})\right) \cdot \exp\left(-\frac{1}{2\eta^2}\boldsymbol{\mathrm{r}}^T\boldsymbol{\mathrm{r}}\right)\\
&\;\;\;\;\;\;\;\;\;\;\;\;\;\;\; \cdot  \exp\left(-\frac{1}{2}\sum_{i=1}^{n}(\boldsymbol{\mathcal{I}}_i- \Mu_i(\Alpha_{\bar{F}}))^T \mathbf{\Sigma}^{-1}_I (\boldsymbol{\mathcal{I}}_i-\Mu_i(\Alpha_{\bar{F}}))\right) \cdot 1_{\mathcal{F}_0}(\boldsymbol{\mathcal{I}}) \cdot 1_{\mathcal{E}_0}(\boldsymbol{\mathrm{y}}, \boldsymbol{\mathrm{r}}).
\end{aligned}
\end{equation*}
 \smallskip
 
\noindent Note, the normalizing constant for the likelihood function derived from the above conditional law is given by
$$\mathbb{P}[\widehat{E}(\mathbf{Y}, \mathbf{R}) = E \;\lvert \;\Beta_{\bar{E}}] \times \mathbb{P}[\widehat{F}_l(\mathbf{I}_l)= F_l \text{ for } l\in \{1,2, \cdots, L\}\;\lvert  \;\Alpha_{F_1},\cdots,\Alpha_{F_L} ].$$
This is because the probability 
$$\mathbb{P}[\widehat{E}(\mathbf{Y}, \mathbf{R}) = E, \; \widehat{F}_l(\mathbf{I}_l)= F_l \text{ for } l\in \{1,2, \cdots, L\}\;\lvert  \; \Beta_{\bar{E}},\;\Alpha_{F_1},\cdots,\Alpha_{F_L} ]$$
decouples due to the independence we assume between the intermediary outcomes and the primary outcome as well as the randomization variable .
Lastly, observe that the probability involving exclusively the auxiliary parameters 
 $$\Alpha_{F_1},\cdots,\Alpha_{F_L}$$
contributes to a constant in our likelihood, a function of $\Beta_{\bar{E}}$.
 Our conditional likelihood is therefore proportional to
$$\left\{\mathbb{P}[\widehat{E}(\mathbf{Y}, \mathbf{R}) = E  \;\lvert \;\Beta_{\bar{E}}] \right\}^{-1}\cdot \rho(\widehat{\Beta}_{\bar{E}}; \Beta_{\bar{E}}, \mathbf{\Sigma}_{\bar{E}}),$$
which agrees with the expression in Proposition \ref{cond:lik}.
\end{proof}

Define the following matrices that we use to characterize the selection event of interest as a union of polyhedral regions in Proposition \ref{polytope}:
$$\mathbf{U}_{\mathbf{s}_E} = \begin{bmatrix} \text{diag}(\mathbf{s}_E) (\mathbf{G}_{E}^T \mathbf{G}_{E} + \epsilon \mathbf{I})^{-1} \mathbf{G}_{E}^T \mathbf{G}_{\bar{E}} \\  \mathbf{G}_{E^c}^T \mathbf{G}_{\bar{E}}- \mathbf{G}_{E^c}^T \mathbf{G}_{E}(\mathbf{G}_{E}^T \mathbf{G}_{E} + \epsilon \mathbf{I})^{-1} \mathbf{G}_{E}^T \mathbf{G}_{\bar{E}}  \\ -\mathbf{G}_{E^c}^T \mathbf{G}_{\bar{E}} +\mathbf{G}_{E^c}^T \mathbf{G}_{E}(\mathbf{G}_{E}^T \mathbf{G}_{E} + \epsilon \mathbf{I})^{-1} \mathbf{G}_{E}^T \mathbf{G}_{\bar{E}} \end{bmatrix},$$
$$\mathbf{V}_{\mathbf{s}_E}= \begin{bmatrix} \text{diag}(\mathbf{s}_E) (\mathbf{G}_{E}^T \mathbf{G}_{E} + \epsilon \mathbf{I})^{-1} & \mathbf{0}  \\   -\mathbf{G}_{E^c}^T \mathbf{G}_{E}(\mathbf{G}_{E}^T \mathbf{G}_{E} + \epsilon \mathbf{I})^{-1}   &  \mathbf{I} \\ \mathbf{G}_{E^c}^T \mathbf{G}_{E}(\mathbf{G}_{E}^T \mathbf{G}_{E} + \epsilon \mathbf{I})^{-1}   &  -\mathbf{I} \end{bmatrix}, \mathbf{W}_{\mathbf{s}_E} = \begin{bmatrix} \text{diag}(\mathbf{s}_E) (\mathbf{G}_{E}^T \mathbf{G}_{E} + \epsilon \mathbf{I})^{-1} & \mathbf{0} \\ - \mathbf{G}_{E^c}^T \mathbf{G}_{E}(\mathbf{G}_{E}^T \mathbf{G}_{E} + \epsilon \mathbf{I})^{-1} & \mathbf{I} \\  \mathbf{G}_{E^c}^T \mathbf{G}_{E}(\mathbf{G}_{E}^T \mathbf{G}_{E} + \epsilon \mathbf{I})^{-1} & -\mathbf{I} \end{bmatrix},$$
\begin{equation*}
\begin{aligned}
\mathbf{t}_{\mathbf{s}_E} &=\begin{bmatrix}\text{diag}(\mathbf{s}_E)(\mathbf{G}_E^T \mathbf{G}_E+\epsilon \mathbf{I})^{-1}\boldsymbol{\Lambda}_E \mathbf{s}_E \\
 -\boldsymbol{\lambda}_{E^c} -  \mathbf{G}_{E^c}^T \mathbf{G}_{E}(\mathbf{G}_{E}^T \mathbf{G}_{E} + \epsilon \mathbf{I})^{-1} \boldsymbol{\Lambda}_E \mathbf{s}_E \\ -\boldsymbol{\lambda}_{E^c} +  \mathbf{G}_{E^c}^T \mathbf{G}_{E}(\mathbf{G}_{E}^T \mathbf{G}_{E} + \epsilon \mathbf{I})^{-1} \boldsymbol{\Lambda}_E \mathbf{s}_E \end{bmatrix}.
\end{aligned}
\end{equation*}
\medskip

\begin{proof}  \ \ Proposition \ref{polytope}. \ \ 
To see a proof for Proposition  \ref{polytope}, the selection of active variables $E$ with signs $\mathbf{s}_E$ is equivalent to 
\begin{equation*}
\text{sign}( \widehat{\Beta}_{E}^{\;\text{LASSO}}) = \mathbf{s}_E\; , \; \|\mathbf{z}\|_{\infty}< 1,
\end{equation*}
where $\widehat{\Beta}_{E}^{\;\text{LASSO}}$ and $\mathbf{z}$ are obtained from the stationary equation at the solution of \eqref{second:stage:Lasso}, given by:
\begin{equation}
\label{KKT:app}
\begin{aligned}
& \begin{pmatrix} \boldsymbol{\mathrm{r}}_E^T & \boldsymbol{\mathrm{r}}^T_{E^c} \end{pmatrix}^T + \begin{pmatrix} (\mathbf{G}_E^T \boldsymbol{\mathrm{y}})^T & (\mathbf{G}_{E^c}^T \boldsymbol{\mathrm{y}})^T  \end{pmatrix}^T  \\
&=   \begin{bmatrix} \mathbf{G}_E^T \mathbf{G}_E + \epsilon \cdot \mathbf{I} & \mathbf{G}_E^T \mathbf{G}_{E^c}\end{bmatrix}^T \widehat{\Beta}_{E}^{\;\text{LASSO}}+\begin{pmatrix}  (\mathbf{\Lambda}_E \mathbf{s}_E)^T & (\mathbf{\Lambda}_{E^c} \mathbf{z} )^T\end{pmatrix}^T.
\end{aligned}
\end{equation}

\smallskip
Using the stationary equation \eqref{KKT:app} for realizations $(\boldsymbol{\mathrm{y}},\boldsymbol{\mathrm{r}})$ and based on the decomposition
$\mathbf{G}_{\bar{F}}^T \boldsymbol{\mathrm{y}} = \mathbf{G}_{\bar{F}}^T \mathbf{G}_{\bar{E}} \widehat{\Beta}_{\bar{E}} + \widehat\Beta^{\perp},$
we first note that the sign constraints are equivalent to
$$\text{diag}(\mathbf{s}_E) (\mathbf{G}_{E}^T \mathbf{G}_{E} + \epsilon \mathbf{I})^{-1} (\mathbf{r}_E + \mathbf{G}_E^T \mathbf{G}_{\bar{E}}\widehat{\Beta}_{\bar{E}} +\widehat\Beta^{\perp}_E -\boldsymbol{\Lambda}_E \mathbf{s}_E)>0.$$
Next observe the inactive coordinates of the stationary equation \eqref{KKT}  yield the equality:
$$\boldsymbol{\mathrm{r}}_{E^c} + \mathbf{G}_{E^c}^T \mathbf{G}_{\bar{E}} \widehat{\Beta}_{\bar{E}} +  \widehat\Beta^{\perp}_{E^c} - \mathbf{G}_{E^c}^T \mathbf{G}_{E}\widehat{\Beta}_{E}^{\;\text{LASSO}} =\boldsymbol{\Lambda}_{E^c} \mathbf{z}.$$
Coupled with the equation 
$$\widehat{\Beta}_{E}^{\;\text{LASSO}} = (\mathbf{G}_{E}^T \mathbf{G}_{E} + \epsilon \mathbf{I})^{-1}(\boldsymbol{\mathrm{r}}_E + \mathbf{G}_E^T \mathbf{G}_{\bar{E}}\widehat{\Beta}_{\bar{E}} +\widehat\Beta^{\perp} -\boldsymbol{\Lambda}_E \mathbf{s}_E),$$
the restriction on the $\ell_\infty$-norm of inactive subgradient vector $\mathbf{z}$ is equivalent to the following constraints
\begin{equation*}
\begin{aligned}
& \boldsymbol{\mathrm{r}}_{E^c} - \mathbf{G}_{E^c}^T \mathbf{G}_{E}(\mathbf{G}_{E}^T \mathbf{G}_{E} + \epsilon \mathbf{I})^{-1} \boldsymbol{\mathrm{r}}_E + \left(\mathbf{G}_{E^c}^T \mathbf{G}_{\bar{E}}- \mathbf{G}_{E^c}^T \mathbf{G}_{E}(\mathbf{G}_{E}^T \mathbf{G}_{E} + \epsilon \mathbf{I})^{-1} \mathbf{G}_{E}^T \mathbf{G}_{\bar{E}}  \right)\widehat{\Beta}_{\bar{E}}\\
&+ \widehat\Beta^{\perp}_{E^c} -   \mathbf{G}_{E^c}^T \mathbf{G}_{E}(\mathbf{G}_{E}^T \mathbf{G}_{E} + \epsilon \mathbf{I})^{-1} \widehat\Beta^{\perp}_{E}>   -\boldsymbol{\lambda}_{E^c} -  \mathbf{G}_{E^c}^T \mathbf{G}_{E}(\mathbf{G}_{E}^T \mathbf{G}_{E} + \epsilon \mathbf{I})^{-1} \boldsymbol{\Lambda}_E \mathbf{s}_E,
\end{aligned}
\end{equation*}
and 
\begin{equation*}
\begin{aligned}
& -\boldsymbol{\mathrm{r}}_{E^c} + \mathbf{G}_{E^c}^T \mathbf{G}_{E}(\mathbf{G}_{E}^T \mathbf{G}_{E} + \epsilon \mathbf{I})^{-1} \boldsymbol{\mathrm{r}}_E - \left(\mathbf{G}_{E^c}^T \mathbf{G}_{\bar{E}}- \mathbf{G}_{E^c}^T \mathbf{G}_{E}(\mathbf{G}_{E}^T \mathbf{G}_{E} + \epsilon \mathbf{I})^{-1} \mathbf{G}_{E}^T \mathbf{G}_{\bar{E}}  \right)\widehat{\Beta}_{\bar{E}}\\
&- \widehat\Beta^{\perp}_{E^c} +   \mathbf{G}_{E^c}^T \mathbf{G}_{E}(\mathbf{G}_{E}^T \mathbf{G}_{E} + \epsilon \mathbf{I})^{-1} \widehat\Beta^{\perp}_{E}>  -\boldsymbol{\lambda}_{E^c} +  \mathbf{G}_{E^c}^T \mathbf{G}_{E}(\mathbf{G}_{E}^T \mathbf{G}_{E} + \epsilon \mathbf{I})^{-1} \boldsymbol{\Lambda}_E \mathbf{s}_E.
\end{aligned}
\end{equation*}
\smallskip

The union of polytopes now follows from considering all possible signs for the active coefficients, which varies over the set $\{-1, 1\}^{|E|}$.
\end{proof}

Fixing some matrices to derive Theorem \ref{correction:factor:cond}, we let 
$$\mathbf{Q} = \begin{bmatrix} \mathbf{G}_E^T \mathbf{G}_E + \epsilon \cdot \mathbf{I} & \mathbf{G}_E^T \mathbf{G}_{E^c}\end{bmatrix}^T, \ \mathbf{S}= -\mathbf{G}_{\bar{F}}^T \mathbf{G}_{\bar{E}}, \ \ \mathbf{s}= \begin{pmatrix}(\boldsymbol{\Lambda}_E \mathbf{s}_E)^T & (\boldsymbol{\Lambda}_{E^c} \mathbf{z})^T \end{pmatrix}^T-\widehat\Beta^{\perp}$$
$$\mathbf{P}= -(\mathbf{Q}^T \mathbf{Q})^{-1}\mathbf{Q}^T \mathbf{S},\ \mathbf{o}=-(\mathbf{Q}^T \mathbf{Q})^{-1}\mathbf{Q}^T \mathbf{s}.$$
Then, we set the following notations based on the above matrices:
$$\boldsymbol{\Theta}_{\bar{E}} = \left(\boldsymbol{\Sigma}^{-1}_{\bar{E}} + \eta^{-2} \mathbf{S}^T \mathbf{S} - \eta^{-2} \mathbf{P}^T \mathbf{Q}^T \mathbf{Q} \mathbf{P}\right)^{-1};$$
$$\mathbf{K}= \boldsymbol{\Theta}_{\bar{E}}\boldsymbol{\Sigma}^{-1}_{\bar{E}},\  \mathbf{l}= \boldsymbol{\Theta}_{\bar{E}}(\eta^{-2} \mathbf{P}^T \mathbf{Q}^T \mathbf{Q} \mathbf{o} - \eta^{-2} \mathbf{S}^T  \mathbf{s}).$$
\smallskip

\begin{proof} \ \ Theorem \ref{correction:factor:cond}. \ \ 
For any fixed set $\bar{E}$, the independence between $\widehat{\Beta}_{\bar{E}}$, $\widehat\Beta^{\perp}$ and $\boldsymbol{\mathrm{r}}$
leads us to their unconditional likelihood:
\begin{equation*}
\begin{aligned}
& \rho(\widehat{\Beta}_{\bar{E}}; \Beta_{\bar{E}}, \boldsymbol\Sigma_{\bar{E}}) \cdot \rho(\boldsymbol{\mathrm{r}}; \mathbf{0}, \eta^2 \mathbf{I}) \cdot \rho(\widehat\Beta^{\perp}; \mathbf{0}, \boldsymbol\Sigma^{\perp}),
\end{aligned}
\end{equation*}
where $\boldsymbol\Sigma^{\perp}$ denotes the covariance for $\widehat\Beta^{\perp}$.
To derive an expression for the conditional likelihood, 
we use a change of measure: $$\mathbf{R} \to \begin{pmatrix} \widehat{\mathbf{B}}^{\text{LASSO}}, & \widehat{\mathbf{Z}}\end{pmatrix}$$ defined through the stationary equation of the randomized LASSO:
 
\begin{equation}
\begin{aligned}
\mathbf{R} &= -\mathbf{G}_{\bar{F}}^T \mathbf{G}_{\bar{E}}\widehat{\Beta}_{\bar{E}} -\widehat\Beta^{\perp} + \begin{bmatrix} \mathbf{G}_E^T \mathbf{G}_E +\epsilon \mathbf{I} \\ \mathbf{G}_{E^c}^T \mathbf{G}_E \end{bmatrix} \widehat{\mathbf{B}}^{\text{LASSO}}  + \begin{pmatrix}  \boldsymbol{\Lambda}_E \mathbf{s}_E\\  \boldsymbol{\Lambda}_{E^c} \widehat{\mathbf{Z}} \end{pmatrix}
\label{CoV}
\end{aligned}
\end{equation}
such that $\widehat{\mathbf{B}}^{\text{LASSO}}$ denotes the active LASSO solution and $\widehat{\mathbf{Z}}$ denotes the subgradient from the $\ell_1$ penalty evaluated at the  inactive coordinates.
This mapping coincides with \eqref{KKT} for our observed data, when $\widehat{\mathbf{B}}^{\text{LASSO}}$ and $\widehat{\mathbf{Z}}$ assume the values $\widehat{\Beta}_{E}^{\;\text{LASSO}}$ and $\mathbf{z}$ respectively.
 Applying the change of measure \eqref{CoV}, the likelihood for the new variables $\widehat{\Beta}_{\bar{E}}$, $\widehat\Beta^{\perp}$, $\widehat{\Beta}_{E}^{\;\text{LASSO}}$ and $\mathbf{z}$ agrees with 
 $$\rho(\widehat{\Beta}_{\bar{E}}; \Beta_{\bar{E}}, \boldsymbol\Sigma_{\bar{E}}) \cdot \rho(\mathbf{S}\widehat{\Beta}_{\bar{E}} + \mathbf{Q}  \widehat{\Beta}_{E}^{\;\text{LASSO}} + \mathbf{s}; 0, \eta^2 \mathbf{I}) \cdot \rho(\widehat\Beta^{\perp}; \mathbf{0}, \boldsymbol\Sigma^{\perp})$$
up to constants.
We use the fact that the Jacobian for this change of measure is free of $\widehat{\mathbf{B}}^{\text{LASSO}}$ and $\widehat{\mathbf{Z}}$ and therefore dissolves as a constant.
Conditioning upon the event 
$$\{\widehat{E}=E, \text{sign}(\widehat{\mathbf{B}}^{\text{\normalfont LASSO}}) = \mathbf{s}_E, \widehat{\mathbf{Z}}= \mathbf{z}, \widehat{\mathbf{B}}^{\perp}=  \widehat\Beta^{\perp}\},$$
observed after solving \eqref{second:stage:Lasso},
yields us the following truncated likelihood for the variables $\widehat{\Beta}_{\bar{E}}$ and $\widehat{\Beta}_{E}^{\;\text{LASSO}}$:
\begin{equation*}
\begin{aligned}
& \left(\int \rho(b; \Beta_{\bar{E}}, \boldsymbol\Sigma_{\bar{E}}) \cdot \rho(\mathbf{S}b + \mathbf{Q} w + \mathbf{s}; \mathbf{0}, \eta^2 \mathbf{I}) \cdot \rho(\widehat\Beta^{\perp}; \mathbf{0}, \boldsymbol\Sigma^{\perp}) \cdot 1_{\left\{\text{sign}(w) = \mathbf{s}_E\right\}}dwdb\right)^{-1}\\
&\times \rho(\widehat{\Beta}_{\bar{E}}; \Beta_{\bar{E}}, \boldsymbol\Sigma_{\bar{E}}) \cdot \rho(\mathbf{S}\widehat{\Beta}_{\bar{E}} + \mathbf{Q}  \widehat{\Beta}_{E}^{\;\text{LASSO}} + \mathbf{s}; \mathbf{0}, \eta^2 \mathbf{I}) \cdot \rho(\widehat\Beta^{\perp}; \mathbf{0}, \boldsymbol\Sigma^{\perp}) \cdot 1_{\left\{\text{sign}( \widehat{\Beta}_{E}^{\;\text{LASSO}}) = \mathbf{s}_E\right\}}\\
&= \left(\int \rho(b; \Beta_{\bar{E}}, \boldsymbol\Sigma_{\bar{E}}) \cdot \rho(\mathbf{S}b + \mathbf{Q} w + \mathbf{s}; \mathbf{0}, \eta^2 \mathbf{I}) \cdot 1_{\left\{\text{sign}(w) = \mathbf{s}_E\right\}}dwdb\right)^{-1}\\
&\;\;\;\;\;\;\;\;\;\;\;\;\;\;\;\;\;\;\;\times \rho(\widehat{\Beta}_{\bar{E}}; \Beta_{\bar{E}}, \boldsymbol\Sigma_{\bar{E}}) \cdot \rho(\mathbf{S}\widehat{\Beta}_{\bar{E}} + \mathbf{Q}  \widehat{\Beta}_{E}^{\;\text{LASSO}} + \mathbf{s}; \mathbf{0}, \eta^2 \mathbf{I}) \;\;\;\cdot 1_{\left\{\text{sign}( \widehat{\Beta}_{E}^{\;\text{LASSO}}) = \mathbf{s}_E\right\}}.
\end{aligned}
\end{equation*}
Integrating out $\widehat{\Beta}_{E}^{\;\text{LASSO}}$ from the likelihood, we are left with the marginal likelihood for $\widehat\Beta_{\bar{E}}$ which is proportional to:
\begin{equation*}
\begin{aligned}
& \left(\int_{\text{sign}(w)= \mathbf{s}_E} \rho(b; \mathbf{K}\Beta_{\bar{E}} +\mathbf{l}, \boldsymbol{\Theta}_{\bar{E}})\cdot \rho(w; \mathbf{P}b+\mathbf{o}, \eta^{-2}\mathbf{Q}^T \mathbf{Q}) dw db\right)^{-1}\\
&\;\;\;\;\;\;\;\;\;\;\;\;\;\;\;\;\;\;\;\;\;\;\;\;\;\;\;\;\;\;\;\;\;\;\;\;\;\;\;\;\;\;\;\;\;\;\;\;\;\;\;\;\;\;\;\;\;\;\;\;\;\;\;\;\;\;\;\;\;\;\;\;\;\;\;\;\;\;\;\;\times\rho(\widehat{\Beta}_{\bar{E}}; \mathbf{K}\Beta_{\bar{E}} +\mathbf{l}, \boldsymbol{\Theta}_{\bar{E}}).
\end{aligned}
\end{equation*}
This completes the derivation of our conditional likelihood.
\end{proof}

\begin{proof} \ \ Theorem \ref{reparam:map}. \ \ 
$\mathrm{(i)}$. We begin by computing the Jacobian associated with the reparameterization mapping; this Jacobian is equal to:
\begin{equation}
\label{Jacobian:equation}
\begin{aligned}
&  \mathbf{K}^{-1}\left(\mathbf{I} + \eta^{-2}\boldsymbol{\Theta}_{\bar{E}} \mathbf{P}^T \mathbf{Q}^T \mathbf{Q} \mathbf{P}  \right) - \eta^{-2} \mathbf{K}^{-1}\boldsymbol{\Theta}_{\bar{E}} \mathbf{P}^T \mathbf{Q}^T \mathbf{Q} \frac{\partial }{\partial \Zeta_{\bar{E}}}(\mathbf{w}^*(\Zeta_{\bar{E}})).
\end{aligned}
\end{equation}
In order to compute $\dfrac{\partial }{\partial \Zeta_{\bar{E}}}(\mathbf{w}^*(\Zeta_{\bar{E}}))$, we note that 
$\mathbf{w}^*(\Zeta_{\bar{E}})$ satisfies the estimating equation:
\begin{equation*}
\eta^{-2} \mathbf{Q}^T \mathbf{Q} \left( \mathbf{w}^*(\Zeta_{\bar{E}}) - \mathbf{P} \Zeta_{\bar{E}} -\mathbf{o}\right) + \grad \text{Barr}_{\mathbf{s}_E}(\mathbf{w}^*(\Zeta_{\bar{E}})) =0. 
\end{equation*}
Taking a derivative of the estimating equation with respect to $\Zeta_{\bar{E}}$ yields us:
\begin{equation*}
\begin{aligned}
& \left(\eta^{-2} \mathbf{Q}^T \mathbf{Q}  + \grad^2 \text{Barr}_{\mathbf{s}_E}(\mathbf{w}^*(\Zeta_{\bar{E}}))\right)  \frac{\partial }{\partial \Zeta_{\bar{E}}}(\mathbf{w}^*(\Zeta_{\bar{E}}))=  \eta^{-2} \mathbf{Q}^T \mathbf{Q} \mathbf{P},
\end{aligned}
\end{equation*}
which implies
$$\frac{\partial }{\partial \Zeta_{\bar{E}}}(\mathbf{w}^*(\Zeta_{\bar{E}})) = \left(\eta^{-2} \mathbf{Q}^T \mathbf{Q}  + \grad^2 \text{Barr}_{\mathbf{s}_E}(\mathbf{w}^*(\Zeta_{\bar{E}}))\right)^{-1} \eta^{-2} \mathbf{Q}^T \mathbf{Q} \mathbf{P}.$$
Plugging the value of $\dfrac{\partial }{\partial \Zeta_{\bar{E}}}(\mathbf{w}^*(\Zeta_{\bar{E}}))$ into \eqref{Jacobian:equation}, the Jacobian assumes the expression
\begin{equation*}
\begin{aligned}
& \mathbf{K}^{-1}\left(\mathbf{I} + \eta^{-2}\boldsymbol{\Theta}_{\bar{E}} \mathbf{P}^T \mathbf{Q}^T \mathbf{Q} \mathbf{P} \right)- \eta^{-4}\mathbf{K}^{-1}\boldsymbol{\Theta}_{\bar{E}} \mathbf{P}^T \mathbf{Q}^T \mathbf{Q}\\
&\;\;\;\;\;\;\;\;\;\;\;\;\;\;\;\;\;\;\;\;\;\;\;\;\;\;\;\;\;\;\;\;\;\;\;\;\;\;\;\;\;\;\;\;\;\;\;\;\;\;\;\;\;\;\;\;\left( \eta^{-2}\mathbf{Q}^T \mathbf{Q} + \grad^2 \text{Barr}_{\mathbf{s}_E} (\mathbf{w}^*(\Zeta_{\bar{E}}))\right)^{-1} \mathbf{Q}^T \mathbf{Q} \mathbf{P}.
\end{aligned}
\end{equation*}
\smallskip

\noindent Using the reparameterization mapping where 
$$\mathbf{K}\Beta_{\bar{E}} + \mathbf{l}= \Psi(\Zeta_{\bar{E}}),$$
the log-posterior is given by:
$$\log\widetilde{\pi}( \Zeta_{\bar{E}} \lvert \widehat{\Beta}_{\bar{E}}) = \log|\boldsymbol{\mathcal{J}}(\Zeta_{\bar{E}})| +\log \pi( \mathbf{K}^{-1}\Psi(\Zeta_{\bar{E}})-\mathbf{K}^{-1}\mathbf{l} \ \lvert \ \widehat{\Beta}_{\bar{E}})$$
where $\pi(\cdot \lvert \ \widehat{\Beta}_{\bar{E}})$ is the working version of the selection-aware posterior in \eqref{approx:posterior}.
Letting $\mathbf{b}^*$, $\mathbf{w}^*$ be the optimal solutions for the problem:
\begin{equation*}
\begin{aligned}
U^*(\Psi(\Zeta_{\bar{E}})) &= \underset{{b, w}}{\inf} \; \Big\{\frac{1}{2}(b-\Psi(\Zeta_{\bar{E}}))^T \boldsymbol{\Theta}_{\bar{E}}^{-1}(b-\Psi(\Zeta_{\bar{E}})) \\
&\;\;\;\;\;\;\;\;\;\;\;\;\;\;+ \frac{1}{2\eta^2}(w-  \mathbf{P} b - \mathbf{o})^T \mathbf{Q}^T \mathbf{Q} (w- \mathbf{P} b- \mathbf{o})+ \text{Barr}_{\mathbf{s}_E}(w)\Big\},
\end{aligned}
\end{equation*}
\smallskip

\noindent we observe that $\log \widetilde{\pi}( \Zeta_{\bar{E}} \lvert \widehat{\Beta}_{\bar{E}})$ is given by
\begin{equation}
\label{post:transformation}
\begin{aligned}
& \log  |\text{det}(\boldsymbol{\mathcal{J}}(\Zeta_{\bar{E}}))|  + \widehat{\Beta}_{\bar{E}}^T \boldsymbol{\Theta}^{-1}_{\bar{E}}  \Psi(\Zeta_{\bar{E}}) - \frac{1}{2} \Psi(\Zeta_{\bar{E}})^T  \boldsymbol{\Theta}^{-1}_{\bar{E}}  \Psi(\Zeta_{\bar{E}}) + U^*(\Psi(\Zeta_{\bar{E}}))\\
&\;\;\;\;\;\;\;\;\;\;\;\;\;\;\;\;\;\;\;\;\;\;\;\;\;\;\;\;\;\;\;\;\;\;\;\;\;\;\;\;\;\;\;\;\;\;\;\;\;\;\;\;\;\;+ \log \pi(\mathbf{K}^{-1}\Psi(\Zeta_{\bar{E}})-\mathbf{K}^{-1}\mathbf{l}) \\
&= \log  |\text{det}(\boldsymbol{\mathcal{J}}(\Zeta_{\bar{E}}))| + \frac{1}{2}{ \mathbf{b}^*}^T \boldsymbol{\Theta}_{\bar{E}}^{-1} \mathbf{b}^*  + \widehat{\Beta}_{\bar{E}}^T \boldsymbol{\Theta}^{-1}_{\bar{E}}  \Psi(\Zeta_{\bar{E}})-  {\mathbf{b}^*}^T \boldsymbol{\Theta}_{\bar{E}}^{-1}\Psi(\Zeta_{\bar{E}}) \\
&\;\;\;\;\;\;\;\;\;\;\;+\frac{1}{2\eta^2 }( \mathbf{w}^*-  \mathbf{P}  \mathbf{b}^* + \mathbf{o})^T \mathbf{Q}^T \mathbf{Q} ( \mathbf{w}^*- \mathbf{P}  \mathbf{b}^* + \mathbf{o})+ \text{Barr}_{\mathbf{s}_E}( \mathbf{w}^*) \\
&\;\;\;\;\;\;\;\;\;\;\;\;\;\;\;\;\;\;\;\;\;\;\;\;\;\;\;\;\;\;\;\;\;\;\;\;\;\;\;\;\;\;\;\;\;\;\;\;\;\;+ \log \pi(\mathbf{K}^{-1}\Psi(\Zeta_{\bar{E}})-\mathbf{K}^{-1}\mathbf{l}).
\end{aligned}
\end{equation}

\noindent Our claim then follows by observing:
\begin{equation*}
\begin{aligned}
  \Psi(\Zeta_{\bar{E}}) &=  \mathbf{b}^* + \eta^{-2} \boldsymbol{\Theta}_{\bar{E}}\left(\dfrac{\partial}{\partial b}  \mathbf{w}^*(b)\Big\lvert_{ \mathbf{b}^*} -\mathbf{P}\right)^T \mathbf{Q}^T \mathbf{Q}  ( \mathbf{w}^*( \mathbf{b}^*)-  \mathbf{P}  \mathbf{b}^* + \mathbf{o})\\
&\;\;\;+ \boldsymbol{\Theta}_{\bar{E}}  \left(\dfrac{\partial}{\partial b}  \mathbf{w}^*(b)\Big\lvert_{ \mathbf{b}^*}\right)^T \grad \text{Barr}_{\mathbf{s}_E}( \mathbf{w}^*(b))\\
&=    \mathbf{b}^* - \eta^{-2} \boldsymbol{\Theta}_{\bar{E}} \mathbf{P}^T \mathbf{Q}^T \mathbf{Q} ( \mathbf{w}^*( \mathbf{b}^*)-  \mathbf{P}  \mathbf{b}^* + \mathbf{o})\\
&=  \mathbf{b}^* + \eta^{-2}\boldsymbol{\Theta}_{\bar{E}}\mathbf{P}^T \mathbf{Q}^T \mathbf{Q} \mathbf{P}  \mathbf{b}^* +  \eta^{-2}\boldsymbol{\Theta}_{\bar{E}} \mathbf{P}^T \mathbf{Q}^T \mathbf{Q} (\mathbf{o}- \mathbf{w}^*( \mathbf{b}^*)),
\end{aligned}
\end{equation*}
where we use the below optimality conditions for $ \mathbf{w}^*$ to deduce the second equality
$$\eta^{-2} \mathbf{Q}^T \mathbf{Q}  ( \mathbf{w}^*( \mathbf{b}^*)-  \mathbf{P}  \mathbf{b}^* + \mathbf{o})+ \grad \text{Barr}_{\mathbf{s}_E}( \mathbf{w}^*( \mathbf{b}^*)) = 0.$$
From the definition of $\Psi(\Zeta_{\bar{E}})$, we have $$ \mathbf{b}^* = \Zeta_{\bar{E}}.$$
The expression for
$\log \widetilde{\pi}( \Zeta_{\bar{E}} \lvert \widehat{\Beta}_{\bar{E}})$ then immediately follows from \eqref{post:transformation}.
\smallskip

\noindent $\mathrm{(ii)}$. Observe that $\grad \log\widetilde{\pi}(\Zeta_{\bar{E}} \lvert \widehat{\Beta}_{\bar{E}})$ equals
\begin{equation*}
\begin{aligned}
&  (\boldsymbol{\mathcal{J}}(\Zeta_{\bar{E}}))^T \cdot \left(\grad \log \pi(\mathbf{K}^{-1}\Psi(\Zeta_{\bar{E}})-\mathbf{K}^{-1}\mathbf{l}) +  \mathbf{K}^T\boldsymbol{\Theta}^{-1}_{\bar{E}}(   \widehat{\Beta}_{\bar{E}} - \Zeta_{\bar{E}})\right)\\
&\;\; - \boldsymbol{\Theta}^{-1}_{\bar{E}} \Psi(\Zeta_{\bar{E}})  +  \boldsymbol{\Theta}^{-1}_{\bar{E}}  \Zeta_{\bar{E}} - \eta^{-2} \mathbf{P}^T \mathbf{Q}^T \mathbf{Q}(\mathbf{w}^*(\Zeta_{\bar{E}}) -\mathbf{P}\Zeta_{\bar{E}} - \mathbf{o})+ \grad  \log |\text{det}(\boldsymbol{\mathcal{J}}(\Zeta_{\bar{E}}))| \\
&=  (\boldsymbol{\mathcal{J}}(\Zeta_{\bar{E}}))^T  \left(\grad\pi(\mathbf{K}^{-1}\Psi(\Zeta_{\bar{E}})-\mathbf{K}^{-1}\mathbf{l})+ \mathbf{K}^T\boldsymbol{\Theta}^{-1}_{\bar{E}}(   \widehat{\Beta}_{\bar{E}} - \Zeta_{\bar{E}})\right) + \grad  \log |\text{det}(\boldsymbol{\mathcal{J}}(\Zeta_{\bar{E}}))| .
\end{aligned}
\end{equation*}
In obtaining the first expression, we use the estimating equation for the solution of the optimization problem \eqref{opt:update}:
$$
\eta^{-2} \mathbf{Q}^T \mathbf{Q} \left( \mathbf{w}^*(\Zeta_{\bar{E}}) - \mathbf{P} \Zeta_{\bar{E}} -\mathbf{o}\right) + \grad \text{Barr}_{\mathbf{s}_E}(\mathbf{w}^*(\Zeta_{\bar{E}})) =0,
$$
and that
$$\dfrac{\partial}{\partial \Zeta_{\bar{E}}} (\Psi(\Zeta_{\bar{E}}))=\left(\mathbf{I} + \eta^{-2}\boldsymbol{\Theta}_{\bar{E}} \mathbf{P}^T \mathbf{Q}^T \mathbf{Q} \mathbf{P}  \right) - \eta^{-2} \boldsymbol{\Theta}_{\bar{E}} \mathbf{P}^T \mathbf{Q}^T \mathbf{Q} \frac{\partial }{\partial \Zeta_{\bar{E}}}(\mathbf{w}^*(\Zeta_{\bar{E}}))=\mathbf{K} \boldsymbol{\mathcal{J}}(\Zeta_{\bar{E}}).$$
The second equality uses the definition of the reparameterization map \eqref{reparam}, that is:
$$ \boldsymbol{\Theta}^{-1}_{\bar{E}}\Zeta_{\bar{E}} - \eta^{-2} \mathbf{P}^T \mathbf{Q}^T \mathbf{Q}(\mathbf{w}^*(\Zeta_{\bar{E}}) -\mathbf{P}\Zeta_{\bar{E}} - \mathbf{o})-  \boldsymbol{\Theta}^{-1}_{\bar{E}}\Psi(\Zeta_{\bar{E}}) =0.$$
The proof is complete by calculating $\grad \log |\text{det}(\boldsymbol{\mathcal{J}}(\Zeta_{\bar{E}}))|$. 
Let us the diagonal matrices
$$\text{\normalfont{diag}}(\grad^3 \text{\normalfont  Barr}_{1;\mathbf{s}_{E;1}}(w_1^*(\Zeta_{\bar{E}})), \cdots, \grad^3  \text{\normalfont  Barr}_{|E|;\mathbf{s}_{E;|E|}} (w_E^*(\Zeta_{\bar{E}}))$$
and
$ \text{\normalfont{diag}}(\eta^{-2}[ \mathbf{N}^{-1} \mathbf{Q}^T \mathbf{Q} \mathbf{P}]_{j})$
by $\mathbf{D}_1$ and $\mathbf{D}_2$ respectively.
Then, the $j$-th coordinate of this gradient vector is equal to:
\begin{equation*}
\begin{aligned}
& \text{Trace}\Bigg( \eta^{-4}\boldsymbol{\mathcal{J}}^{-1}(\Zeta_{\bar{E}})  \mathbf{K}^{-1}\boldsymbol\Theta_{\bar{E}}  \mathbf{P}^T \mathbf{Q}^T \mathbf{Q} \mathbf{N}^{-1}\left(\dfrac{\partial }{\partial \zeta_{j,\bar{E}}} (\mathbf{N}(\Zeta_{\bar{E}}))\right) \mathbf{N}^{-1} \mathbf{Q}^T \mathbf{Q} \mathbf{P}  \Bigg)\\
&\text{Trace}\Bigg( \eta^{-4}\boldsymbol{\mathcal{J}}^{-1}(\Zeta_{\bar{E}})  \mathbf{K}^{-1}\boldsymbol\Theta_{\bar{E}}  \mathbf{P}^T \mathbf{Q}^T \mathbf{Q} \mathbf{N}^{-1}\mathbf{D}_1 \mathbf{D}_2 \mathbf{N}^{-1} \mathbf{Q}^T \mathbf{Q} \mathbf{P}  \Big),
\end{aligned}
\end{equation*}
which completes the proof of our claim.

\end{proof}

\subsection{Working version of selection-aware posterior}
\label{section:2:Supplement}

For the sake of completeness, we provide the probabilistic motivation behind the working selection-aware posterior \eqref{approx:posterior} based on a Laplace approximation for a multivariate Gaussian integral.
The next Proposition derives the approximation for $I(\Beta_{\bar{E}} )$ as an upper bound on the probability of selection. 
We use $\widehat{\mathbf{B}}_{\bar{E}}$ and $\widehat{\mathbf{B}}^{\text{LASSO}}$ to denote the random variables assuming the instances $\widehat{\Beta}_{\bar{E}}$ and $\widehat{\Beta}_{E}^{\;\text{LASSO}}$ respectively and use $C$ to denote a constant free of $\Beta_{\bar{E}} $.
\begin{Proposition}
\label{approx:posterior_2}
Suppose the joint multivariate Gaussian likelihood for $\widehat{\Beta}_{\bar{E}}$ and $\widehat{\Beta}_{E}^{\;\text{LASSO}}$ is proportional to
$$\rho(\widehat{\Beta}_{\bar{E}}; \mathbf{K}\Beta_{\bar{E}} +\mathbf{l}, \boldsymbol{\Theta}_{\bar{E}})\cdot \rho(\widehat{\Beta}_{E}^{\;\text{LASSO}}; \mathbf{P}b+\mathbf{o}, \eta^{-2}\mathbf{Q}^T \mathbf{Q}).$$
For an arbitrary convex, compact region $K$, we have
the following upper bound for  the probability $\mathbb{P}\left[\begin{pmatrix}\widehat{\mathbf{B}}_{\bar{E}} &  \widehat{\mathbf{B}}^{\text{LASSO}} \end{pmatrix} \in K \;\lvert \;\Beta_{\bar{E}} \right]$:
\begin{align*}
&C\cdot \exp\Big( -\inf_{(b,w)\in K} \Big\{\;\frac{1}{2} (b-\mathbf{K}\Beta_{\bar{E}}-\mathbf{l})^T \boldsymbol{\Theta}_{\bar{E}}^{-1}(b-\mathbf{K}\Beta_{\bar{E}}-\mathbf{l})\\
&\;\;\;\;\;\;\;\;\;\;\;\;\;\;\;\;\;\;\;\;\;\;\;\;\;\;\;\;\;\;\;\;\;\;\;\;\;\;\;\;\;\;\;\;\;\;\;\;+ \frac{1}{2\eta^2}(w- \mathbf{P}b-\mathbf{o})^T \mathbf{Q}^T \mathbf{Q} (w-\mathbf{P}b-\mathbf{o})  \Big\} \Big).
\end{align*}
\end{Proposition}

\begin{proof}
Denoting the MGF of the random variables $\widehat{\mathbf{B}}_{\bar{E}}$ and $\widehat{\mathbf{B}}^{\text{LASSO}}$ at $\boldsymbol{\eta}_1$ and $\boldsymbol{\eta}_2$ as $\Lambda(\boldsymbol{\eta}_1, \boldsymbol{\eta}_2)$ with respect to their Gaussian law,
we observe the following:
\begin{equation*}
\setlength{\jot}{10pt}
\begin{aligned}
& \log \mathbb{P}\left[\begin{pmatrix}\widehat{\mathbf{B}}_{\bar{E}} &  \widehat{\mathbf{B}}^{\text{LASSO}} \end{pmatrix} \in K \;\lvert \;\Beta_{\bar{E}} \right]\\
 &\leq \log\mathbb{E}\left[\exp\left(\boldsymbol{\eta}_1^T \widehat{\mathbf{B}}_{\bar{E}} + \boldsymbol{\eta}_2^T \widehat{\mathbf{B}}^{\text{LASSO}} -\inf_{(b, w)\in K } \{\boldsymbol{\eta}_1^T b + \boldsymbol{\eta}_2^T w\} \right)\; \Big\lvert \;\Beta_{\bar{E}}  \;\right ] \\
&= \sup_{(b, w)\in K} -\boldsymbol{\eta}_1^T b -\boldsymbol{\eta}_2^T w  + \log \Lambda(\boldsymbol{\eta}_1, \boldsymbol{\eta}_2).
\end{aligned}
\end{equation*}

Since the above inequality holds for any $\boldsymbol{\eta}_1, \boldsymbol{\eta}_2$, 
optimizing over the parameters $\boldsymbol{\eta}_1, \boldsymbol{\eta}_2$ and using a minimax equality for a convex and compact subset $K$ yields the following bound on the log-selection probability after ignoring constants:
\begin{equation*}
\begin{aligned}
&-\sup_{\boldsymbol{\eta}_1, \boldsymbol{\eta}_2}\;\;\inf_{(b, w)\in K }\Big\{\boldsymbol{\eta}_1^T b + \boldsymbol{\eta}_2^T w  -  \log \Lambda(\boldsymbol{\eta}_1, \boldsymbol{\eta}_2)\Big\}\\
&= -\inf_{(b, w)\in K}\sup_{\boldsymbol{\eta}_1, \boldsymbol{\eta}_2}\Big\{\boldsymbol{\eta}_1^T b + \boldsymbol{\eta}_2^T w  -  \log \Lambda(\boldsymbol{\eta}_1, \boldsymbol{\eta}_2)\Big\}\\
&= -\inf_{(b, w)\in K} \Big\{\;\frac{1}{2} (b-\mathbf{K}\Beta_{\bar{E}}-\mathbf{l})^T \boldsymbol{\Theta}_{\bar{E}}^{-1}(b-\mathbf{K}\Beta_{\bar{E}}-\mathbf{l}) \\
&\;\;\;\;\;\;\;\;\;\;\;\;\;\;\;\;\;\;\;+ \frac{1}{2\eta^2}(w- \mathbf{P}b-\mathbf{o})^T \mathbf{Q}^T \mathbf{Q} (w-\mathbf{P}b-\mathbf{o})  \Big\}.
\end{aligned}
\end{equation*}
\end{proof}
Notice, the selection probability in Theorem \ref{correction:factor:cond} is calculated over the region $\{(b, w): \text{sign}(w) = s_E\}$, which is convex but not compact. We remark however that the approximate posterior obtained by appending a prior to the conditional likelihood and plugging in the bound in Proposition \ref{approx:posterior_2} works well under a large enough compact subset $K'$ of the selection region, for all $\Beta_{\bar{E}} $, in a bounded set of probability close to $1$ under $\pi(\cdot)$. Rigorous asymptotic justification of the approximate selection-aware posterior is based on consistency guarantees aligned along the moderate deviations scale in \citep{panigrahi2019approximate}; we refer interested readers to this previous work for the asymptotic guarantees behind the Laplace-type approximation.

\subsection{Supplementary details for radiogenomic analysis}
\label{section:3:Supplement}

\subsubsection{Radiogenomic feature construction: Pathway scores}
\label{sec:pathway:scores}

Pathway based methods provide significant benefits by offering interpretability, as gene functions are exerted collectively and may vary based on several factors such as genetic modification, disease state, or environmental stimuli. Using pathways provides an intuitive way and a stable context for assessing the biological activity \citep{hanzelmann2013gsva}. Pathway scores are computed using the gene-set variation analysis (GSVA), which estimates a value per sample for the variation of pathway activity within an entire gene expression set, using a non-paramteric and unsupervised approach \citep{hanzelmann2013gsva}. In other words, a pathway score assesses the relative variability of gene expression of the genes in the pathway as compared to expression of genes not in the pathway. We give a brief overview of the analytical procedure for GSVA next.

   Let $Z$ be the $p \times n$ matrix of normalized gene expression values corresponding to $p$ genes and $n$ samples ($p \gg n$). Let $G = \{g_1,\ldots,g_m \}$ represent a collection of pathways (also referred to as gene-sets). Each pathway $g_k$ is defined as $g_k \subset \{1,\ldots,p \}$ with $|g_k|$ denoting its cardinality. Let the expression profile for the gene $i$ be given as $z_i = (z_{i1},\ldots,z_{in})$.
	
	Firstly, in the context of the sample population distribution GSVA evaluates whether a gene $i$ is highly or lowly expressed in the sample $j$. To compare distinct expression profiles on the same scale, an expression-level statistic is computed. A non-parametric kernel estimation of the cumulative density function is performed for each $z_i$ using a Gaussian kernel, that is, we compute $\hat{F}_{s_i}(z_{ij}) = \frac{1}{n} \sum\limits_{r=1}^n \Phi(\frac{z_{ij}-z_{ir}}{s_i})$. Here $s_i$ is the gene-specific bandwidth parameter controlling the resolution of the kernel estimation. These statistics $\hat{F}_{s_i}(z_{ij})$ are converted to ranks $r_{(i)j}$ for each sample $j$. The ranks $r_{(i)j}$ are normalized further as $t_{ij} = |\frac{p}{2}-r_{(i)j}|$ so that the tails of the rank distribution are up-weighted while computing the enrichment score. The normalized ranks $t_{ij}$ are used to compute a Kolmogorov-Smirnov (KS) type random walk statistic for $l = 1,\ldots,p$ as 
	\begin{equation*}
	\eta_{jk}(l) = \frac{\sum\limits_{i=1}^l |t_{ij}|^\tau I(u_{(i)} \in g_k)}{\sum\limits_{i=1}^p |t_{ij}|^\tau I(u_{(i)} \in g_k)} - \frac{\sum\limits_{i=1}^l I(u_{(i)} \in g_k)}{p - |g_k|}.
	\end{equation*}
	Here $I(u_{(i)} \in g_k)$ is an indicator taking the value $1$ if the gene corresponding to the rank $i$ expression level statistic belongs to the pathway $g_k$ and $\tau$ is the parameter describing the weight of the tail. By identifying if the genes in a pathway are more likely to belong to either tail of the rank distribution, the statistic $\eta_{jk}(l)$ produces a distribution over the genes. The enrichment score for the pathway $g_k$ and the sample $j$ is constructed by converting the corresponding KS-like statistic as $S_{jk} = \max\limits_l (0, \eta_{jk}(l)) - \min\limits_l (0, \eta_{jk}(l)) $. $S_{jk}$ has a clear biological interpretation as it emphasizes genes in pathways that are concordantly activated in one direction only, which are either over-expressed or under-expressed relative to the overall population \citep{hanzelmann2013gsva}. Low enrichment is shown for pathways containing genes strongly acting in both directions.	The computations are performed using the GSVA package in R obtained from the Bioconductor package \citep{gentleman2004bioconductor} under the default settings for the choice of parameters.

 \subsubsection{Radiogenomic feature construction: MRI scans and radiomic phenotypes}\label{sec:mri}
    We consider four types of MRI sequences which include (i) native (T1), (ii) post-contrast T1-weighted (T1Gd), (iii) T2-weighted (T2), and (iv) T2 fluid attenuated inversion recovery (FLAIR) volumes. Each of these sequences display different types of tissues with varying contrasts based on the tissue characteristics. Note that the whole brain MRI scans are three dimensional objects and have an array structure. In Figure \ref{fig: mri}, we show an axial slice from the MRI scan of a LGG subject corresponding to all four imaging sequences with the segmented tumor region indicated by a (red) boundary overlaid on those images. This tumor region is further classified into sub-regions (NC, ED and ET) by GLISTRboost as shown in Figure \ref{fig: mri}.
	
	\begin{figure}[h]
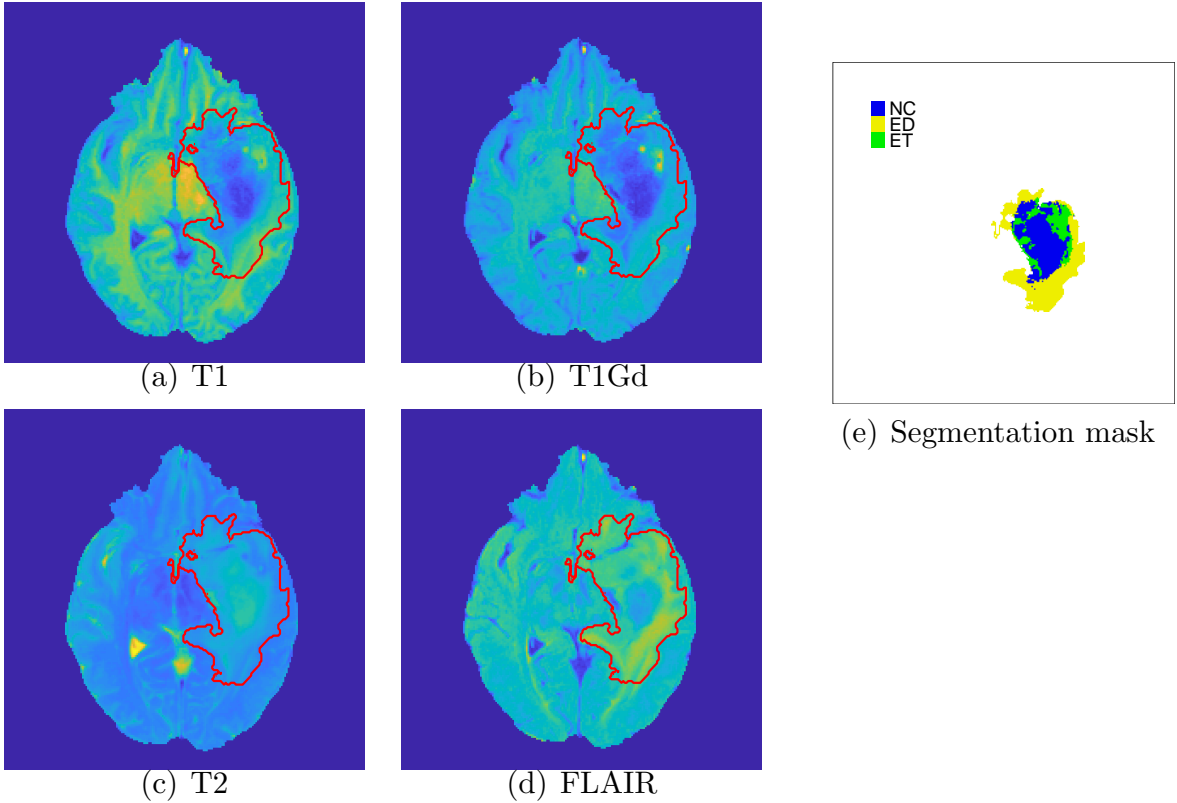

		\centering
		\resizebox{\textwidth}{!}{%
		\begin{tabular}{cc}
		    \begin{tabular}{cc}
			\subfigure[T1]{\includegraphics[trim=1cm 1cm 1cm 2cm, clip, width=1.6in, height=1.6in]{t1.pdf}} & \subfigure[T1Gd]{\includegraphics[trim=1cm 1cm 1cm 2cm, clip, width=1.6in, height=1.6in]{t1Gd.pdf}} \\
			\subfigure[T2]{\includegraphics[trim=1cm 1cm 1cm 2cm, clip, width=1.6in, height=1.6in]{t2.pdf}} & \subfigure[FLAIR]{\includegraphics[trim=1cm 1cm 1cm 2cm, clip, width=1.6in, height=1.6in]{flair.pdf}} \\
		\end{tabular} & \subfigure[Segmentation mask]{\includegraphics[trim=0.05cm 0.05cm 0cm 0cm, clip, width=1.6in, height=1.6in]{mask_labelled.pdf}}
		\end{tabular}
		}
		\caption{An axial slice of a brain MRI from four modalities: T1, T1Gd, T2 and FLAIR for a LGG subject. The segmented tumor region is shown with an (red) overlaid boundary. Segmentation mask indicates the necrotic and non-enhancing tumor core (NC), the peritumoral edema (ED) and the enhancing tumor (ET) regions.}
		\label{fig: mri}
	\end{figure}
	
	Voxel-level features are usually extracted to provide additional insight into the tumor physiology, and have been studied in the context of the progression (or regression) of tumors. Summary statistics such as percentiles, quartiles, skewness, kurtosis etc., are evaluated to represent a region of interest and are used as features/covariates in downstream analysis \citep{baek2012percent,just2014improving,song2013true}. However, some of their drawbacks include the subjectivity in the choice of number and location of summary features, and limitations in terms of capturing entire information from the histogram of intensity values. To address this, we consider the smoothed density arising from the voxel-level intensity histogram which incorporates granular characteristics of tumor heterogeneity \citep{saha2016demarcate}. The variability in these intensity histograms across the subjects is captured through the scores from a principal component analysis on the space of density functions using a Reimannian-geometric framework. In other words, these principal component scores obtained from the density functions act as the radiomic phenotypes which capture the heterogeneity in the tumor voxels from the MRI scans. We include these details in Sections \ref{sec:mri} and \ref{subsec:pc_scores}.

    \subsection{Radiogenomic feature construction:  Principal Component Scores}\label{subsec:pc_scores}
	Next, we discuss the construction of the principal component scores when we have probability density functions (pdfs) as data objects corresponding to $n$ samples. Without loss of generality, let us consider densities on $[0,1]$ and let $\mathcal{F}$ denote the Banach manifold of such pdfs defined as $\mathcal{F} = \{f : \mathbb[0,1] \rightarrow \mathbb{R}_{+} | \int_0^1 f(x) dx = 1 \}$. A non-parametric Fisher-Rao Reimannian metric which is invariant to reparameterizations can be defined, however it is computationally challenging to compute the geodesic paths and distances using this metric \citep{srivastava2016functional}. 
	
	An equivalent representation of the space $\mathcal{F}$ via the square-root transformation (SRT) representation \citep{bhattacharyya1943measure} simplifies computations. The SRT is defined as a function $h = +\sqrt{f}$ (we omit the $+$ sign hereafter for notational convenience). Also, the inverse mapping is unique and is simply given by $f = h^2$ \citep{kurtek2015bayesian}. Space of SRTs is given by $\mathcal{H} = \{h : [0,1] \rightarrow \mathbb{R}^{+} | \int_0^1 h(x)^2 dx = 1 \}$ and represents the postive orthant of a unit Hilbert sphere \citep{lang2012fundamentals}. The $\mathbb{L}^2$ Riemmanian metric on $\mathcal{H}$ can be defined as $\langle\langle \delta h_1 , \delta h_2 \rangle\rangle= \int_0^1 \delta h_1(t)\delta h_2(t) dt$, where $\delta h_1,\delta h_2 \in T_h(\mathcal{H})$ and $T_{h}(\mathcal{H}) = \{ \delta h : [0,1] \rightarrow \mathbb{R} | \int_0^1 h(x)\delta h(x)dx = 0\}$. The geodesic paths and lengths can now be analytically computed due to the Reimannian geometry of $\mathcal{H}$ equipped with the $\mathbb{L}^2$ metric. The geodesic distance between $h_1,h_2\in\mathcal{H}$ is simply given by $d(h_1,h_2)=\theta=\cos^{-1}\big(\int_0^1 h_{1}(x)h_{2}(x) dx\big)$.
	
	The geometry of the space of SRTs can be used to define an average (or mean) density corresponding to a sample of density functions. This allows us to efficiently summarize and visualize a sample of densities. The average pdf can be computed using a generalized version of mean on a metric space called the \textit{Karcher} mean \citep{karcher1977riemannian}. Suppose we have $n$ pdfs $f_1,\ldots,f_n$ and the corresponding SRTs as $h_1,\ldots,h_n$. The sample Karcher mean $\bar{h}$ on $\mathcal{H}$ is defined as the minimizer of the Karcher variance $\rho(\bar{h}) = \sum\limits_{i=1}^n d(h_i,\bar{h})^2_{\mathbb{L}^2}$, that is, $\bar{h} = \text{argmin}_{h \in \mathcal{H}} d(h_i,\bar{h})^2_{\mathbb{L}^2}$. Algorithm \ref{algo: karcher} presents a gradient-based approach to compute the Karcher mean on $\mathcal{H}$ \citep{dryden1998statistical}.
	
	\begin{algorithm}[h]
	\caption{Sample Karcher mean of densities}\label{algo: karcher}
	\begin{algorithmic}[1]
	\State $\bar{h}_0$ (initial estimate for the Karcher mean) $\gets$ any one of the densities in the sample OR the extrinsic average. Set $j \gets 0$ and $\epsilon_1, \epsilon_2 > 0$ be small.
	\State For $i=1,\ldots,n$ compute $u_i = \exp^{-1}_{\bar{h}_j}(h_i)$.
	\State Compute the average direction in the tangent space $\bar{u} = \frac{1}{n} \sum_{i=1}^n u_i$.
	\If {$||\bar{u}||_{L^2} < \epsilon_1$} 
	    \State \Return $\bar{h}_j$ as the Karcher mean.
	\Else {} 
	    \State$\bar{h}_{j+1} = \exp_{\bar{h}_j}(\epsilon_2 \bar{u})$.
	    \State Set $j \gets j+1$.
	    \State Return to step $2$.
	\EndIf
	\end{algorithmic}
	\end{algorithm}

	Note that the Karcher mean of the sample pdfs is an intrinsic average that is computed directly on $\mathcal{H}$ (or equivalently $\mathcal{F}$). Hence we have a mean which is an actual pdf (Karcher mean) and a distance function. Here the inverse exponential map, denoted by $\exp^{-1}_{h_1}: \mathcal{H} \mapsto T_{h_1}(\mathcal{H})$, is given by $\exp^{-1}_{h_1}(h_2) = (\theta/\sin(\theta))(h_2 - h_1\cos(\theta))$. The exponential map at a point $h_1 \in \mathcal{H}$, denoted by $\exp: T_{h_1}(\mathcal{H}) \mapsto \mathcal{H}$, is defined as $ \exp_{h_1}(\delta h) = \cos(\| \delta h \|)h_1+ \sin(\| \delta h \|)(\delta h/\|\delta h\|)$,
	where $\|\delta h\|=\big(\int_0^1 \delta h(x)^2 dx\big)^{1/2}$.
	
	Under the standard settings, visualizing the space of pdfs intuitively is not straight forward. Principal component analysis (PCA) is an effective method to explore the variability in the pdfs through their primary modes of variation in the data. Note that the tangent space is a vector space (Euclidean), hence PCA can be implemented, as in standard problems. Algorithm \ref{algo: pca} describes the computation of PCA on the space generated by the pdfs $f_1,\ldots,f_n$ and the corresponding SRTs $h_1,\ldots,h_n$.
	
	\begin{algorithm}[h]
	\caption{PCA on $\mathcal{P}$}\label{algo: pca}
	\begin{algorithmic}[1]
	\State Compute the Karcher mean of $h_1,\ldots,h_n$ as $\bar{h}$.
	\For {$i=1,\ldots,n$} 
	    \State Compute projections ($v_i = \exp^{-1}_{\bar{h}} (h_i)$) of $h_i$ onto $T_{\bar{h}}(\mathcal{H})$.
	\EndFor
	\State Evaluate sample covariance matrix $K = \frac{1}{n-1} \sum\limits_{i=1}^n v_i v_i^\top \in \mathbb{R}^{m \times m}$.
	\State Compute the SVD of $K = U\Sigma U^\top$.
	\end{algorithmic}
	\end{algorithm}

	Note that the first $r$ columns of $U$ (denoted as $\tilde{U} \in \mathbb{R}^{m \times r}$) span the $r$-dimensional principal subspace. We can compute the principal coefficients as $X = V\tilde{U}$, where $V^\top = [v_1~ v_2~ \ldots~ v_n] \in \mathbb{R}^{m \times n}$. These principal coefficients $X$ (principal component scores) act as Euclidean coordinates corresponding to densities $f_i$ and can be used as predictors for downstream modeling.
	
\begin{figure}[h]
		\centering
		\includegraphics[scale=0.8]{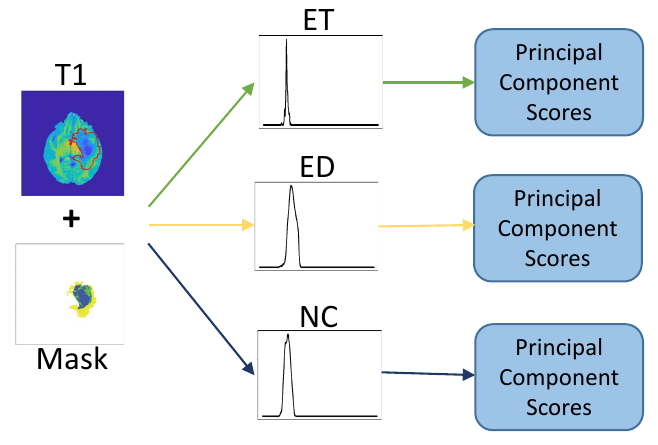}
		\caption{Workflow to obtain the principal component scores from the tumor sub-regions (ND, ET, and ED) in the T1 MRI scan.}
		\label{fig:pc_scores}
	\end{figure}

    In Figure \ref{fig:pc_scores}, we represent the workflow to construct the principal component scores using the tumor intensity values for the T1 MRI sequence. The same workflow is followed for the other three MRI sequences to compute the principal component scores. The number of principal components to include could be chosen using a threshold on the percent variance explained.
    
\end{document}